\documentclass[aps,preprint,floats,epsf,epsfig,nofootinbib,letter]{revtex4}
\usepackage{epsfig}
\usepackage{graphicx}
\usepackage{dcolumn}
\usepackage{bm}
\usepackage{subfigure}
\usepackage{hyperref}                 
\usepackage{amsmath}
\usepackage{xcolor}
\usepackage{soul}

\begin{document}
\def\be{\begin{eqnarray}}
\def\en{\end{eqnarray}}
\def\non{\nonumber}
\def\la{\langle}
\def\ra{\rangle}
\def\ov{\overline}
\def\Br{{\mathcal B}}
\def\A{{\mathcal A}}
\def\B{{\cal B}}
\def\D{{\cal D}}
\def\Bbar{\overline{\cal B}}
\def\bfB{{\rm\bf B}}
\def\bfBB{{\rm\bf B}\overline{\rm\bf B}}
\def\bfBcBc{{\rm\bf B}_c\overline{\rm\bf B}_c}
\def\BB{{{\cal B}_c \overline {\cal B}_c}}
\def\BD{{{\cal B} \overline {\cal D}}}
\def\DB{{{\cal D} \overline {\cal B}}}
\def\DD{{{\cal D} \overline {\cal D}}}
\def\sq{\sqrt}


\title{Two-body charmed anti-charmed baryonic $B$ decays}

\author{Chun-Khiang Chua}
\affiliation{Department of Physics and Center for High Energy Physics,
Chung Yuan Christian University,
Chung-Li, Taiwan 320, Republic of China}

\date{\today}

\begin{abstract}
We study the rates of two-body charmed anti-charmed baryonic $\overline B\to {\cal B}_c \overline {\cal B}_c$ decays using the topological amplitude approach.
All amplitudes of $\overline B\to {\cal B}_c(\bf {\bar 3_f}) \overline {\cal B}_c(\bf { 3_f})$,
${\cal B}_c(\bf 6_f) \overline {\cal B}_c(\bf { 3_f})$,
${\cal B}_c(\bf {\bar 3_f}) \overline {\cal B}_c(\bf {\bar 6_f})$
and
${\cal B}_c(\bf 6_f) \overline {\cal B}_c(\bf {\bar 6_f})$ decays
are decomposed topologically.
SU(3) breaking effects on these amplitudes, depending on the position of the $s$-quark line, are modeled.
Using existing data as inputs, we obtained the following results. 
(i) In the low-lying $\overline B\to {\cal B}_c(\bf {\bar 3_f}) \overline {\cal B}_c(\bf { 3_f})$ decays, we find that
the exchange diagram is sizable.
Furthermore, there is a large cancellation between internal $W$-tree and exchange $W$-tree amplitudes.   
The SU(3) breaking is sizable, 35\% SU(3) breaking effects are needed, and they work differently in different amplitudes.
Namely, the internal $W$-tree amplitude is enlarged, 
while the exchange $W$-tree amplitude is reduced.
The rates of $\overline B\to {\cal B}_c(\bf {\bar 3_f}) \overline {\cal B}_c(\bf { 3_f})$ decays with excited ${\cal B}_c(\bf {\bar 3_f})$, such as 
$\Lambda_c(2595)^+$, $\Xi_c(2790)^{+,0}$, are also studied.
(ii) The $\overline B\to {\cal B}_c(\bf 6_f) \overline {\cal B}_c(\bf { 3_f})$ decays, with low-lying $ \overline {\cal B}_c(\bf { 3_f})$ 
and low-lying and excited ${\cal B}_c(\bf 6_f)$ baryons, such as $\Sigma_c(2520)^{+,0}$, $\Xi_c(2645)^{+,0}$, $\Omega_c(2770)^0$, 
are studied with some predictions on rates obtained.
Furthermore, as the excited states are spin-3/2 states, some of their rates are highly suppressed by the kinematic factor.
(iii) The $\overline B\to {\cal B}_c(\bf {\bar 3_f}) \overline {\cal B}_c(\bf {\bar 6_f})$ decays with low-lying charmed anti-charmed baryons are studied with some predictions on rates obtained.
(iv)~Uncertainties in most predicted rates are large, reflecting our current poor understanding of the related SU(3) breaking effects.
Measuring these rates can provide very useful information about these effects.
\end{abstract}

\pacs{11.30.Hv,  
      13.25.Hw,  
      14.40.Nd}  

\maketitle


\vfill\eject

\section{Introduction}

Recently, there is some progress in two-body charmed anti-charmed baryonic $B$ decays. 
The current situation is summarized in Table~\ref{tab: expt}~\cite{ParticleDataGroup:2024cfk,
Belle:2018kzz,
Belle:2019pze,
Belle:2025nup,
LHCb:2025ueu,
Belle:2019bgi}.
In particular, the 
$B^-\to \Xi_c^+\bar\Sigma_c(2455)^{--}$, $\bar B^0\to \Xi_c^0\bar\Sigma_c(2455)^0 $, 
$\bar B^0\to \Lambda^+_c\bar \Lambda^-_c$ and $\bar B^0_s\to \Lambda^+_c\bar \Lambda^-_c$ decays were reported in 2025
by Belle II \cite{Belle:2025nup} and LHCb \cite{LHCb:2025ueu}, respectively.

We summarize in Table~\ref{tab: quantum number} the quantum numbers of the charmed baryons related to those in Table~\ref{tab: expt}, see \cite{ParticleDataGroup:2024cfk} (and also \cite{Chua:2018lfa}).
One can see that the modes shown in Table~\ref{tab: expt} belong to
$\overline B\to {\cal B}_c(\bf {\bar 3_f}) \overline {\cal B}_c(\bf { 3_f})$,
${\cal B}_c(\bf 6_f) \overline {\cal B}_c(\bf { 3_f})$
and
${\cal B}_c(\bf {\bar 3_f}) \overline {\cal B}_c(\bf {\bar 6_f})$ decays, 
where $\bf {\bar 3_f}$ and $\bf 6_f$ are SU(3) anti-triplet and sextet multiplets, respectively, 
with low-lying and excited charmed anti-charmed baryon final states.
Indeed, we have data on the rates of 4 low-lying $\overline B\to {\cal B}_c(\bf {\bar 3_f}) \overline {\cal B}_c(\bf { 3_f})$ decays,
1 $\overline B\to {\cal B}_c(\bf {\bar 3_f}) \overline {\cal B}_c(\bf { 3_f})$ decay with an excited ${\cal B}_c(\bf {\bar 3_f})$,
1 low-lying and 1 excited $\overline B\to {\cal B}_c(\bf 6_f) \overline {\cal B}_c(\bf { 3_f})$ decays,
and 2 low-lying $\overline B\to {\cal B}_c(\bf {\bar 3_f}) \overline {\cal B}_c(\bf {\bar 6_f})$ decays, totally 9 modes.
Given the recent experimental progress, it is timely and interesting to investigate these decays systematically.

There are some theoretical studies on low-lying $\overline B\to\B_c\overline \B_c$ decays, 
\cite{Chernyak:1990ag, Ball:1990fw,Cheng:2005vd,Chen:2006fsa,Cheng:2009yz,Hsiao:2023mud,Rui:2024xgc,Geng:2025yna}.
Early calculations often yield overly large rates.
As the direct theoretical calculations of the decays of these modes remain challenging, 
we shall use the well established topological amplitude formalism~\cite{Zeppenfeld:1980ex,Chau:tk,Chau:1990ay,Gronau:1994rj,Gronau:1995hn,Chiang:2004nm,Cheng:2014rfa,Savage:ub}
in this study. 
This approach has been applied to baryonic $B$ decays, see, for example, \cite{,Chua:2003it,Chua:2013zga,Chua:2016aqy,Chua:2022wmr}.

The importance of the contribution from the $W$-exchange diagram and the need for SU(3) breaking in low-lying $\overline B\to {\cal B}_c(\bf {\bar 3_f}) \overline {\cal B}_c(\bf { 3_f})$ decays have been reported by LHCb \cite{LHCb:2025ueu}.
In fact, the need for $W$-exchange diagram has been pointed out in ref. \cite{Hsiao:2023mud}, see also ref. \cite{Hsiao:2019wyd},
and the need for SU(3) breaking is also pointed out in ref. \cite{Rui:2024xgc}.
We shall include both internal $W$-tree and $W$-exchange diagrams with modeling of SU(3) breaking effects in these topological amplitudes.
Existing data shown in Table~\ref{tab: expt} will be used as inputs to predict the rates of other modes.
Although data on $\overline B\to {\cal B}_c(\bf 6_f) \overline {\cal B}_c(\bf { 3_f})$ and $ {\cal B}_c(\bf {\bar 3_f}) \overline {\cal B}_c(\bf {\bar 6_f})$ decays are relatively less,
as we shall see, these modes are internal $W$-tree modes. 
It is expected that their amplitudes, for final states within the same multiplets, are highly related.
Hence, even with these data, it is likely that some predictions on rates can still be made.

The layout of this paper is as follows. 
Formalism is given in Sec.~II, 
where topological amplitudes in SU(3) limit and modeling of SU(3) breaking effects are introduced.
The resulting decompositions of all $\overline B\to {\cal B}_c \overline {\cal B}_c$ decay amplitudes and formulas of rates are given in Sec. III.
In Sec. IV, numerical results on rates of $\overline B\to {\cal B}_c(\bf {\bar 3_f}) \overline {\cal B}_c(\bf { 3_f})$,
${\cal B}_c(\bf 6_f) \overline {\cal B}_c(\bf { 3_f})$
and
${\cal B}_c(\bf {\bar 3_f}) \overline {\cal B}_c(\bf {\bar 6_f})$ decays with low-lying and some excited final states are shown. 
We give our conclusions in Sec. V, which is followed by an appendix.

\begin{table}[t!]
\caption{\label{tab: expt}
Experimental results of $\bar B_{u,d,s}\to\bfBcBc$ branching ratios in the unit of $10^{-4}$. 
The upper limits are at 90\% confidence level.
}
\begin{ruledtabular}
\begin{tabular}{lcccr}
Mode
          & LHCb
          & Belle/ Belle-II
          & PDG~\cite{ParticleDataGroup:2024cfk}
          \\
\hline $B^-\to\Xi_c^0\bar \Lambda_c^-$
          & 
          & $9.51\pm2.10\pm0.88$ \cite{Belle:2018kzz}
          & $9.5\pm2.3$
          \\
$B^-\to\Xi_c^{\prime 0}\bar \Lambda_c^-$
          & 
          & $3.4\pm 2.0$ \cite{Belle:2019pze}
          & $<6.5$
          \\
$B^-\to\Xi_c^{0}(2645)\bar \Lambda_c^-$
          & 
          & $4.4\pm 2.4$ \cite{Belle:2019pze}
          & $<7.9$
          \\
$B^-\to\Xi_c^{0}(2790)\bar \Lambda_c^-$
          & 
          & $1.1\pm 0.4$ \cite{Belle:2019pze}
          & $1.1\pm 0.4$
          \\
$B^-\to \Xi_c^+\bar\Sigma_c(2455)^{--} $
          & 
          & $5.74\pm1.11\pm0.42^{+2.47}_{-1.53}$ \cite{Belle:2025nup}
          & 
          \\          
\hline $\bar B^0\to\Xi_c^+\bar \Lambda_c^-$
          & 
          & $11.6\pm4.2\pm1.5$ \cite{Belle:2019bgi}
          & $11\pm 8$
          \\
$\bar B^0\to \Lambda^+_c\bar \Lambda^-_c$
          & $0.101^{+0.027}_{-0.028}\pm0.008\pm0.015$ \cite{LHCb:2025ueu}
          &  
          & $<0.16$
          \\
$\bar B^0\to \Xi_c^0\bar\Sigma_c(2455)^0 $
          & 
          & $4.83\pm1.12\pm0.37^{+0.72}_{-0.60}$ \cite{Belle:2025nup}
          & 
          \\        
\hline $\bar B^0_s\to \Lambda^+_c\bar \Lambda^-_c$
          & $0.50\pm0.13\pm0.05\pm0.08$   \cite{LHCb:2025ueu}
          &
          & $<0.8$
          \\                                                         
\end{tabular}
\end{ruledtabular}
\end{table}

\begin{table}[t!]
\caption{
The quantum numbers of charmed baryons involved in this study, see \cite{ParticleDataGroup:2024cfk} (and also \cite{Chua:2018lfa}).}
\label{tab: quantum number}
\begin{ruledtabular}
\begin{tabular}{lcc}
    $\B_c$
    & SU(3) multiplet
    & $J^P$
    \\
    \hline
    $\Lambda^+_c$, $\Xi_c^{+,0}$
    & ${\bf \bar 3_f}$
    & $\frac{1}{2}^+$
     \\
     $\Lambda_c(2595)^+$, $\Xi_c(2790)^{+,0}$
    & ${\bf \bar 3_f}$
    & $\frac{1}{2}^-$
    \\  
\hline      
    $\Sigma_c(2455)^{++,+,0}$, $\Xi_c^{\prime +,0}$, $\Omega^0_c$
    & ${\bf 6_f}$ 
    & $\frac{1}{2}^+$
    \\
     $\Sigma_c(2520)^{++,+,0}$, $\Xi_c(2645)^{+,0}$, $\Omega_c(2770)^0$
    & ${\bf 6_f}$
    & $\frac{3}{2}^+$
    \\
\end{tabular}
\end{ruledtabular}
\end{table}

\section{Formalism}

We follow \cite{Chua:2003it, Chua:2025xsg} to obtain the decomposition of decay amplitudes in terms of topological amplitudes.
We consider the SU(3) symmetric case first and include the modeling of SU(3) breaking effects later.

\begin{figure}[t]
\centering
 \subfigure[]{
  \includegraphics[width=0.45\textwidth]{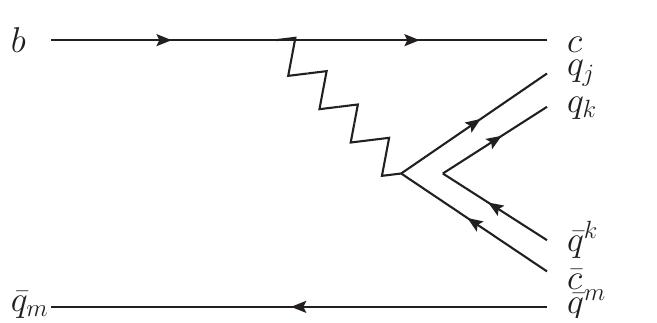}
}
\hspace{12pt}
\subfigure[]{
  \includegraphics[width=0.45\textwidth]{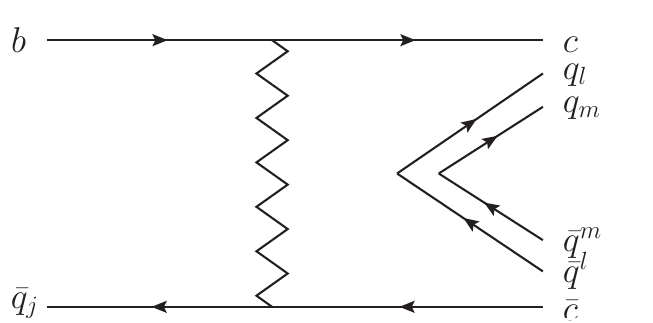}
}
\caption{Topological diagrams of 
  (a) $C$ (internal $W$-tree) and (b) $E$ ($W$-exchange)
  amplitudes in $\overline B$ to charmed baryon pair decays. 
  These are flavor flow diagrams. 
} \label{fig:TA}
\end{figure}

\subsection{SU(3) symmetric case}
 
For low-lying sextet ($\bf 6_f$) charmed baryons, 
we have
\be
&&
\quad
\B_c^{\{11\}}=\sqrt2 \Sigma^{++}_c,\,\,
\B_c^{\{12\}}=\B_c^{\{21\}}=\Sigma^{+}_c,\,\,
\B_c^{\{22\}}=\sqrt2 \Sigma^0_c,
\non\\
&&
\B_c^{\{13\}}=\B_c^{\{31\}}=\Xi^{\prime +}_c,\,\,
\B_c^{\{23\}}=\B_c^{\{32\}}=\Xi^{\prime 0}_c,\,\,
\B_c^{\{33\}}=\sqrt2 \Omega^0_c,
\en
while for low-lying anti-triplet ${\bf \bar 3_f}$ charmed baryons, we have
\be
\B_c^{[12]}=-\B_c^{[21]}=\Lambda^+_c,\,\,
\B_c^{[23]}=-\B_c^{[32]}=\Xi^0_c,\,\,
\B_c^{[31]}=-\B_c^{[13]}=\Xi^+_c.
\en
For simplicity, we only show the assignments for low-lying charmed baryons in the above equations.
One can readily obtain similar assignments for excited charmed baryons based on their SU(3) quantum numbers.

The $\overline B \to \B_c \overline \B_c $ decays is governed by a $(\bar s c)_{V-A}(\bar c b)_{V-A}$ operator.
The flavor structure of the above operator can be expressed as $(\bar q_j H^j c)\,(\bar c b)$ with $q^j=(u,\,d,\,s)$ and
\begin{equation}
H= \left(
\begin{array}{c}
0 \\
0 \\
1 
\end{array}
\right).
\label{eq: Hj}
\end{equation}
Consequently, the Hamiltonians governing various $\overline B \to \B_c \overline \B_c $ decays are given by
\be
H_{\rm eff}\left(\overline B \to \B_c(\bf {\bar 3_f}) \overline \B_c(\bf { 3_f}) \right)
&=& 
       C_1 \, \overline B_m H^j \overline \B_{c [jk]}  \B_c^{[km]}
      +\frac{1}{2}E_1 \, \overline B_j H^j \overline \B_{c [lm]}  \B_c^{[ml]},
 \non\\
H_{\rm eff}\left(\overline B \to \B_c(\bf 6_f) \overline \B_c(\bf { 3_f}) \right)
&=& 
       C_2 \, \overline B_m H^j \overline \B_{c\{jk\}}  \B_c^{[km]},
\non\\
H_{\rm eff}\left(\overline B \to \B_c(\bf {\bar 3_f}) \overline \B_c(\bf {\bar 6_f}) \right)
&=& 
       C_3 \, \overline B_m H^j \overline \B_{c[jk]}  \B_c^{\{km\}},
 \non\\
H_{\rm eff}\left(\overline B \to \B_c(\bf 6_f) \overline \B_c(\bf {\bar 6_f}) \right)
&=& 
       C_4 \, \overline B_m H^j \overline \B_{c\{jk\}}  \B_c^{\{km\}}
      +\frac{1}{2}E_4 \, \overline B_j H^j \overline \B_{c\{lm\}}  \B_c^{\{ml\}}.
\label{eq: H0}
\en
These $C$ and $E$ amplitudes denote internal $W$-tree and $W$-exchange amplitudes, respectively, see Fig. \ref{fig:TA}.
Note that following refs. \cite{Hsiao:2023mud, Rui:2024xgc}, we do not include penguin amplitudes, as their contributions to averaged rates are expected to be subleading and can be neglected.
We have two topological amplitudes in $\overline B \to \B_c(\bf {\bar 3_f}) \overline \B_c(\bf { 3_f})$ decays,
another two topological amplitudes in $\overline B \to \B_c(\bf 6_f) \overline \B_c(\bf {\bar 6_f})$ decays,
but only need one topological amplitude each in $\overline B \to \B_c(\bf 6_f) \overline \B_c(\bf { 3_f})$ or $\overline B \to \B_c(\bf {\bar 3_f}) \overline \B_c(\bf {\bar 6_f})$ decays.

The number of topological amplitudes can be easily understood as follows.
Recall that one has the following SU(3) decompositions:
\be
\bar{\bf 3}\otimes{\bf 3}&=&{\bf 8}\oplus{\bf 1},
\non\\
{\bf 6}\otimes{\bf 3}&=&{\bf 8}\oplus{\bf 10},
\non\\
\bar{\bf 3}\otimes\bar{\bf 6}&=&{\bf 8}\oplus\overline{\bf 10},
\non\\
{\bf 6}\otimes\bar{\bf 6}&=&{\bf 27}\oplus{\bf 8}\oplus{\bf 1}.
\en
The decaying $\overline B$ meson and the tree operator form a product of ${\bf 3}\otimes\bar{\bf 3}={\bf 8}\oplus{\bf 1}$, 
while the final states $\B_c(\bf {\bar 3_f}) \overline \B_c(\bf { 3_f})$ and $\B_c(\bf 6_f) \overline \B_c(\bf {\bar 6_f})$ each also have an ${\bf 8}$ and a ${\bf 1}$, 
which can match the ${\bf 8}$ and the ${\bf 1}$ in the combination of the decaying $\overline B$ meson and the tree operator, giving two independent amplitudes in each case.
This is, however, not the case in the other two decays. 
Although $\B_c(\bf 6_f) \overline \B_c(\bf { 3_f})$ and $\B_c(\bf {\bar 3_f}) \overline \B_c(\bf {\bar 6_f})$ final states each have an ${\bf 8}$ that can match to the ${\bf 8}$ in the decaying $\overline B$ meson and the tree operator, they do not have any ${\bf 1}$, resulting only one independent amplitude in each case.

The above effective Hamiltonians are for the $\Delta S=-1$ transition, and they can be easily transformed into those for the $\Delta S=0$ transition, 
by simply replacing $H$ in Eq. (\ref{eq: Hj}) with
\begin{equation}
H'= \left(
\begin{array}{c}
0 \\
1 \\
0 
\end{array}
\right),
\label{eq: Hj'}
\end{equation}
and with $C_i$ and $E_i$ replaced by $C'_i$ and $E'_i$, respectively. 
It is understood that these amplitudes are related by the Cabibbo-Kobayashi-Maskawa (CKM) matrix elements, namely
\be
C^{\prime}_i= \frac{V_{cb} V^*_{cd}}{V_{cb} V^*_{cs}} C_i,
\quad
E^{\prime}_i= \frac{V_{cb} V^*_{cd}}{V_{cb} V^*_{cs}} E_i.
\en

\subsection{Including SU(3) breaking}

To model the SU(3) breaking in $\overline B\to \bar \B_c \B_c$ decays, we define
\begin{equation}
M= \left(
\begin{array}{ccc}
0 &0 &0\\
0 &0 &0 \\
0 &0 &1
\end{array}
\right),
\label{eq: M}
\end{equation}
which can generate SU(3) breaking terms in amplitudes, as the breaking originates from the relatively large $s$-quark mass.
Including the SU(3) breaking parts, the Hamiltonians for the $\Delta S=-1$ transition now become
\be
H_{\rm eff}\left(\overline B \to \B_c(\bf {\bar 3_f}) \overline \B_c(\bf { 3_f}) \right)
&=& 
       C_1 \, \overline B_m H^j \overline \B_{c [jk]}  \B_c^{[km]}
       +\delta_v C_1 \, \overline B_m H^j M_j^{j'}\overline \B_{c [j'k]}  \B_c^{[km]}
 \non\\
 &&+ \delta_s     C_1 \, \overline B_{m'} M^{m'}_m H^j \overline \B_{c [jk]}  \B_c^{[km]}
      + \delta_c     C_1 \, \overline B_m  H^j \overline \B_{c [jk']} M^{k'}_k \B_c^{[km]}
\non\\      
     && +\frac{1}{2}E_1 \, \overline B_j H^j \overline \B_{c [lm]}  \B_c^{[ml]}
     +\frac{1}{2}\delta_v E_1 \, \overline B_{j'} M^{j'}_j H^j \overline \B_{c [lm]}  \B_c^{[ml]}
\non\\
 &&+\frac{1}{2} \delta_c E_1 \, \overline B_j H^j (\overline \B_{c [l'm]}  M^{l'}_l \B_c^{[ml]}+\overline \B_{c [lm']}  M^{m'}_m \B_c^{[ml]}),
 \non\\
H_{\rm eff}\left(\overline B \to \B_c(\bf 6_f) \overline \B_c(\bf { 3_f}) \right)
&=& 
       C_2 \, \overline B_m H^j \overline \B_{c\{jk\}}  \B_c^{[km]}       
       +\delta_v C_2 \, \overline B_m H^j M_j^{j'}\overline \B_{c \{j'k\}}  \B_c^{[km]}
 \non\\
 &&+ \delta_s     C_2 \, \overline B_{m'} M^{m'}_m H^j \overline \B_{c \{jk\}}  \B_c^{[km]}
      + \delta_c     C_2 \, \overline B_m  H^j \overline \B_{c \{jk'\}} M^{k'}_k \B_c^{[km]},
\non\\    
H_{\rm eff}\left(\overline B \to \B_c(\bf {\bar 3_f}) \overline \B_c(\bf {\bar 6_f}) \right)
&=& 
       C_3 \, \overline B_m H^j \overline \B_{c[jk]}  \B_c^{\{km\}}
              +\delta_v C_3 \, \overline B_m H^j M_j^{j'}\overline \B_{c [j'k]}  \B_c^{\{km\}}
 \non\\
 &&+ \delta_s     C_3 \, \overline B_{m'} M^{m'}_m H^j \overline \B_{c [jk]}  \B_c^{\{km\}}
      + \delta_c     C_3 \, \overline B_m  H^j \overline \B_{c [jk']} M^{k'}_k \B_c^{\{km\}},
 \non\\
H_{\rm eff}\left(\overline B \to \B_c(\bf 6_f) \overline \B_c(\bf {\bar 6_f}) \right)
&=& 
       C_4 \, \overline B_m H^j \overline \B_{c\{jk\}}  \B_c^{\{km\}}
              +\delta_v C_4 \, \overline B_m H^j M_j^{j'}\overline \B_{c \{j'k\}}  \B_c^{\{km\}}
 \non\\
 &&+ \delta_s     C_4 \, \overline B_{m'} M^{m'}_m H^j \overline \B_{c \{jk\}}  \B_c^{\{km\}}
      + \delta_c     C_4 \, \overline B_m  H^j \overline \B_{c \{jk'\}} M^{k'}_k \B_c^{\{km\}}
\non\\    
   &&   +\frac{1}{2}E_4 \, \overline B_j H^j \overline \B_{c\{lm\}}  \B_c^{\{ml\}}
      +\frac{1}{2}\delta_v E_4 \, \overline B_{j'} M^{j'}_j H^j \overline \B_{c \{lm\}}  \B_c^{\{ml\}}
\non\\
 &&+\frac{1}{2} \delta_c E_4 \, \overline B_j H^j 
        (\overline \B_{c \{l'm\}}  M^{l'}_l \B_c^{\{ml\}}+\overline \B_{c \{lm'\}}  M^{m'}_m \B_c^{\{ml\}}),
\label{eq: H1}
\en
and the Hamiltonians for the $\Delta S=0$ transition can be obtained readily.

\begin{figure}[t]
\centering
 \subfigure[]{
  \includegraphics[width=0.45\textwidth]{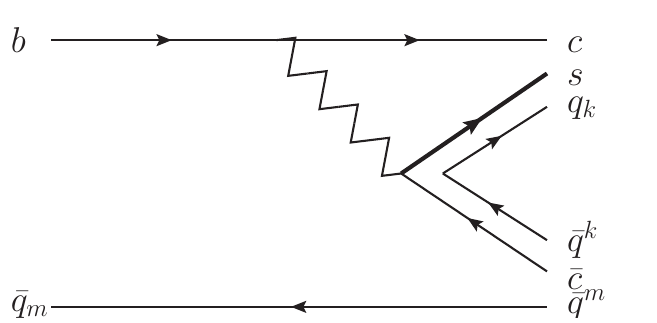}
}
\hspace{12pt}
\subfigure[]{
  \includegraphics[width=0.45\textwidth]{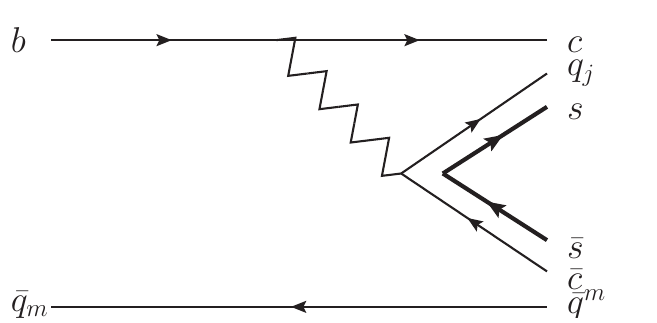}
}
\\\subfigure[]{
  \includegraphics[width=0.45\textwidth]{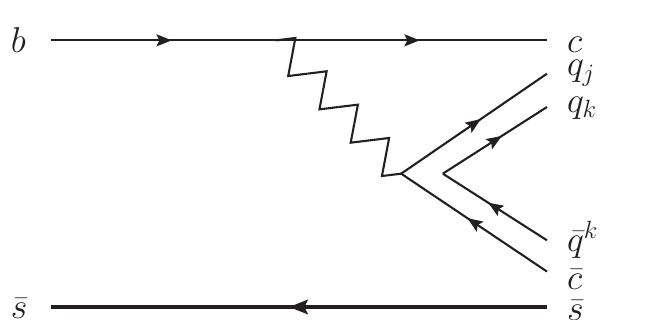}
}
\hspace{12pt}
\subfigure[]{
  \includegraphics[width=0.45\textwidth]{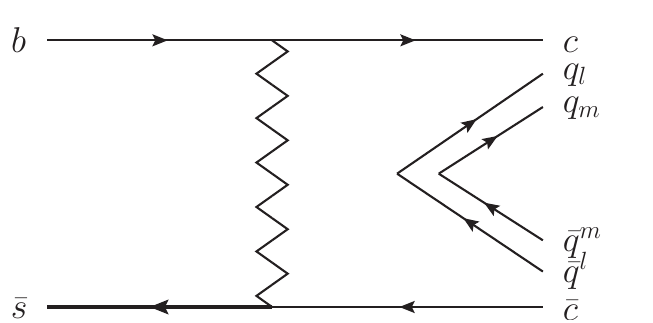}
}
\\\subfigure[]{
  \includegraphics[width=0.45\textwidth]{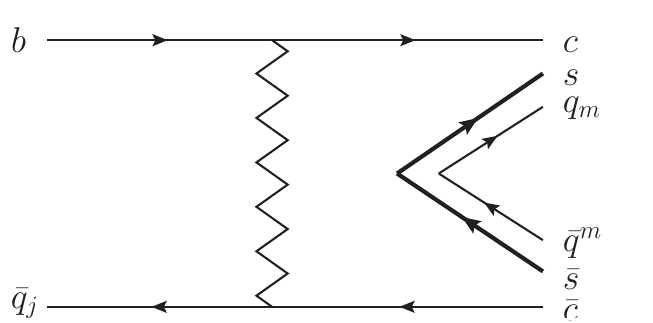}
}
\hspace{12pt}
\subfigure[]{
  \includegraphics[width=0.45\textwidth]{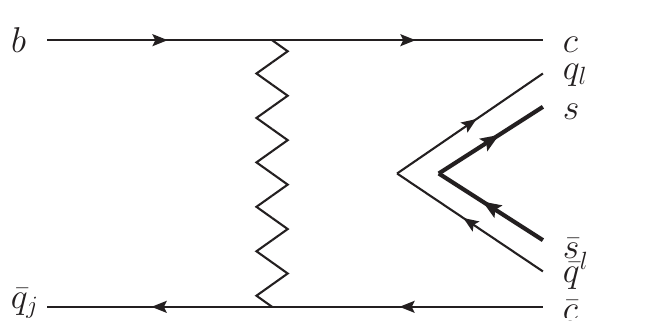}
}
\caption{Internal $W$-tree diagrams containing SU(3) breaking effects,  in
  (a) vertex line, (b) pair creation line, and (c) spectator line;  
  $W$-exchange diagrams containing SU(3) breaking effects, in (d) vertex line and (e)-(f) pair creation lines.
} \label{fig:TA s}
\end{figure}

To simplify the notation, we define
\be
C^{(\prime)}_{i,v}&\equiv& C^{(\prime)}_i+\delta_v C^{(\prime)}_i\equiv(1+\delta_v^{c_i}) C^{(\prime)}_i,
\non\\
C^{(\prime)}_{i,s}&\equiv& C^{(\prime)}_i+\delta_s C^{(\prime)}_i\equiv(1+\delta_s^{c_i}) C^{(\prime)}_i,
\non\\
C^{(\prime)}_{i,c}&\equiv& C^{(\prime)}_i+\delta_c C^{(\prime)}_i\equiv(1+\delta_c^{c_i}) C^{(\prime)}_i,
\non\\
C^{(\prime)}_{i,vs}&\equiv& C^{(\prime)}_i+\delta_v C^{(\prime)}_i+\delta_s C^{(\prime)}_i
\equiv(1+\delta_v^{c_i}+\delta_s^{c_i}) C^{(\prime)}_i,
\non\\
C^{(\prime)}_{i,cs}&\equiv& C^{(\prime)}_i+\delta_c C^{(\prime)}_i+\delta_s C^{(\prime)}_i
\equiv(1+\delta_c^{c_i}+\delta_s^{c_i}) C^{(\prime)}_i,
\non\\
C^{(\prime)}_{i,vcs}&\equiv& C^{(\prime)}_i+\delta_v C^{(\prime)}_i+\delta_c C^{(\prime)}_i+\delta_s C^{(\prime)}_i
\equiv(1+\delta_v^{c_i}+\delta_c^{c_i}+\delta_s^{c_i}) C^{(\prime)}_i,
\label{eq: C SU(3) X}
\en
for $i=1-4$, and, likewise,
\be
E^{(\prime)}_{j,v}&\equiv& E^{(\prime)}_{j}+\delta_v E^{(\prime)}_{j}\equiv(1+\delta_v^{e_j}) E^{(\prime)}_{j},
\non\\
E^{(\prime)}_{j,c}&\equiv& E^{(\prime)}_{j}+\delta_c E^{(\prime)}_{j}\equiv(1+\delta_c^{e_j}) E^{(\prime)}_{j},
\non\\
E^{(\prime)}_{j,cc}&\equiv& E^{(\prime)}_{j}+2\delta_c E^{(\prime)}_{j}\equiv(1+2\delta_c^{e_j}) E^{(\prime)}_{j},
\non\\
E^{(\prime)}_{j,vc}&\equiv& E^{(\prime)}_{j}+\delta_v E^{(\prime)}_{j}+\delta_c E^{(\prime)}_{j}
\equiv(1+\delta_v^{e_j}+\delta_c^{e_j}) E^{(\prime)}_j,
\non\\
E^{(\prime)}_{j,vcc}&\equiv& E^{(\prime)}_{j}+\delta_v E^{(\prime)}_{j}+2\delta_c E^{(\prime)}_{j}
\equiv(1+\delta_v^{e_j}+2\delta_c^{e_j}) E^{(\prime)}_j,
\label{eq: E SU(3) X}
\en
for $j=1, 4$.
These $\delta_v^{c_i}$, $\delta_s^{c_i}$, $\delta_c^{c_i}$, $\delta_v^{e_i}$ and $\delta_c^{e_i}$ parameters should be the measures of the degree of the SU(3) breaking in amplitudes.
As we shall see using Eq. (\ref{eq: H1}), $\overline B\to\B_c\bar\B_c$ decay amplitudes can be decomposed into the above amplitudes, in addition to $C^{(\prime)}_i$ and $E^{(\prime)}_i$.

We depict in Fig. \ref{fig:TA s} some topological diagrams containing SU(3) breaking effects.
Internal $W$-tree diagrams containing SU(3) breaking effects,  in (a)~vertex line, (b)~pair creation line, and (c)~spectator line,  
and $W$-exchange diagrams containing SU(3) breaking effects, in (d)~vertex line and (e)-(f) pair creation lines, are shown in the figure.
When other quark lines are $u$, $d$ quark lines, these diagrams can be identified as $C_{i,v}$, $C_{i,c}$, $C_{i,s}$,  $E_{i,v}$ and $E_{i,c}$, 
where the subscripts $v$, $c$ and $s$, stand for vertex, pair creation and spectator, respectively, see also Eq. (\ref{eq: C SU(3) X}).
For example, we have
\be
C_{i,v}=(1+\delta_v^{c_i}) C_i,
\quad
C_{i,s}=(1+\delta_s^{c_i}) C_i,
\en
where the SU(3) breaking parameters $\delta_v^{c_i}$ and $\delta_s^{c_i}$ occur in $C_{i,v}$ and $C_{i,s}$, respectively,
as the $s$-quark lines are in the vertex and spectator of the internal $W$-tree diagrams, respectively, as shown Fig.~\ref{fig:TA s} (a) and (c),
respectively.

Note that there may be more than one $s$-quark line in these diagrams. 
For example, from Eq. (\ref{eq: C SU(3) X}), we have
\be
C_{i,vs}=(1+\delta_v^{c_i}+\delta_s^{c_i}) C_i.
\en
It can be understood as follows. 
In $C_{i,vs}$, the $s$ quarks can be in the vertex and the spectator at the same time.
Hence, we have both SU(3) breaking parameters $\delta_v^{c_i}$ and $\delta_s^{c_i}$ occurring in $C_{i,vs}$ at the same time. 
 
 \section{Topological amplitudes and branching ratios}

\subsection{Topological amplitudes}

Decay amplitudes of various $\overline B_q\to \B_c\overline \B_c$ decays in $\Delta S=-1$ and  $\Delta S=0$ transitions, in terms of topological amplitudes, can be readily obtained using Eq. (\ref{eq: H1}).
We give the decompositions of $\overline B_q\to \B_c({\bf \bar 3_f})\overline \B_c ({\bf 3_f})$ decay amplitudes in Table \ref{tab: BtoBcBcbar I},
$\overline B_q\to \B_c({\bf 6_f})\overline \B_c ({\bf 3_f})$ decay amplitudes in Table \ref{tab: BtoBcBcbar II},
$\overline B_q\to \B_c({\bf \bar 3_f})\overline \B_c ({\bf \bar 6_f})$ decay amplitudes in Table \ref{tab: BtoBcBcbar III},
and $\overline B_q\to \B_c({\bf 6_f})\overline \B_c ({\bf \bar 6_f})$ decay amplitudes in Table \ref{tab: BtoBcBcbar IV}.
Note that although some of the decay modes in $\overline B_q\to \B_c({\bf 6_f})\overline \B_c ({\bf \bar 6_f})$ decays are kinematically forbidden, their amplitudes are also listed for completeness.
In fact, the $\B_c({\bf 6_f})$ or $\bar\B_c({\bf \bar 6_f})$, in a kinematically forbidden two-body decay, 
may participate in a multi-body $\overline B_q$ decay virtually. 
Hence, the two-body decay amplitude can still be useful.

Note that the decompositions of decay amplitudes in these tables are not only valid for low-lying $\B_c({\bf \bar 3_f})$ and $\B_c({\bf 6_f})$ states.
They are also applicable for excited $\B_c$ and/or $\overline \B_c$ final states.
For example, from Table \ref{tab: quantum number}, we see that $\Xi_c^{0}(2645)$ is in a ${\bf 6_f}$ multiplet.
Hence, $B^-\to\Xi_c^{0}(2645)\bar \Lambda_c^-$ decay is also a $\overline B_q\to \B_c({\bf 6_f})\overline \B_c ({\bf 3_f})$ decay 
with amplitude decomposition similar to those given in Table~\ref{tab: BtoBcBcbar II}.
Indeed, by referring to Table~\ref{tab: BtoBcBcbar II}, we should have $A(B^-\to\Xi_c^{0}(2645)\bar \Lambda_c^-)=-\tilde C_{2,v}$, 
where a different notation, $\tilde C_{2,v}$, is used to differentiate it from the $C_{2,v}$ of the low-lying states.

\begin{table}[t!]
\caption{\label{tab: BtoBcBcbar I}
$\overline B_q\to \B_c({\bf \bar 3_f})\overline \B_c ({\bf 3_f})$ decay amplitudes in $\Delta S=-1$ and  $\Delta S=0$ transitions.}
\footnotesize{
\begin{ruledtabular}
\begin{tabular}{llcllccccr}
Mode
          & $A \left(\overline B_q\to \B_c({\bf \bar 3_f})\overline \B_c ({\bf 3_f})\right)$
          & Mode
          & $A \left(\overline B_q\to \B_c({\bf \bar 3_f})\overline \B_c ({\bf 3_f})\right)$
          \\
\hline
$B^-\to \Xi_c^0 \bar \Lambda_c^{-}$
          & $C_{1,v}$      
          & $\overline B{}^0\to \Xi_c^+ \bar \Lambda_c^{-}$
          & $C_{1,v}$
          \\
$\overline B{}_s^0\to \Xi_c^0 \bar \Xi_c^{0}$
          & $-C_{1,vs}-E_{1,vc}$
          & $\overline B{}_s^0\to \Xi_c^+ \bar \Xi_c^{-}$
          & $-C_{1,vs}-E_{1,vc}$
          \\
$\overline B{}_s^0\to \Lambda_c^+ \bar \Lambda_c^{-}$
          & $-E_{1,v}$
          \\
\hline
$B^-\to \Xi_c^0 \bar \Xi_c^{-}$
          & $C'_{1,c}$
          & $\overline B{}^0\to \Lambda_c^+ \bar \Lambda_c^{-}$
          & $-C'_1-E'_1$
          \\          
          $\overline B{}^0\to \Xi_c^0 \bar \Xi_c^{0}$
          & $-C'_{1,c}-E'_{1,c}$ 
          & $\overline B{}^0\to \Xi_c^+ \bar \Xi_c^{-}$
          & $-E'_{1,c}$         
          \\ 
          $\overline B{}_s^0\to \Lambda_c^+ \bar \Xi_c^{-}$
          & $C'_{1,s}$
          \\                                            
\end{tabular}
\end{ruledtabular}
}
\end{table}

\begin{table}[t!]
\caption{\label{tab: BtoBcBcbar II}
$\overline B_q\to \B_c({\bf 6_f})\overline \B_c ({\bf 3_f})$ decay amplitudes in $\Delta S=-1$ and  $\Delta S=0$ transitions.}
\footnotesize{
\begin{ruledtabular}
\begin{tabular}{lclc}
Mode
          & $A \left(\overline B_q\to \B_c({\bf 6_f})\overline \B_c ({\bf 3_f})\right)$
          & Mode
          & $A \left(\overline B_q\to \B_c({\bf 6_f})\overline \B_c ({\bf 3_f})\right)$
          \\
\hline
           $B^-\to \Xi_c^{\prime 0} \bar \Lambda_c^{-}$
          & $-C_{2,v}$ 
          & $B^-\to \Omega_c^{0} \bar \Xi_c^{-}$
          & $\sqrt2 C_{2,vc}$ 
          \\     
          $\overline B{}^0\to \Xi_c^{\prime +} \bar \Lambda_c^{-}$
          & $C_{2,v}$
          & $\overline B{}^0\to \Omega_c^{0} \bar \Xi_c^{0}$
          & $-\sqrt2 C_{2,vc}$
          \\
          $\overline B{}_s^0\to \Xi_c^{\prime +} \bar \Xi_c^{-}$
          & $-C_{2,vs}$
          & $\overline B{}_s^0\to \Xi_c^{\prime 0} \bar \Xi_c^{0}$
          & $C_{2,vs}$
           \\
\hline
         $B^-\to \Sigma_c^0 \bar \Lambda_c^{-}$
          & $-\sqrt2 C'_2$
          & $B^-\to \Xi_c^{\prime 0} \bar \Xi_c^{-}$
          & $C'_{2,c}$
         \\
         $\overline B{}^0\to \Sigma_c^+ \bar \Lambda_c^{-}$
          & $C'_2$        
          & $\overline B{}^0\to \Xi_c^{\prime 0} \bar \Xi_c^{0}$
          & $-C'_{2,c}$ 
          \\
          $\overline B{}_s^0\to \Sigma_c^+ \bar \Xi_c^{-}$
          & $-C'_{2,s}$
          & $\overline B{}_s^0\to \Sigma_c^0 \bar \Xi_c^{0}$
          & $\sqrt2 C'_{2,s}$
          \\                                            
\end{tabular}
\end{ruledtabular}
}
\end{table}

\begin{table}[t!]
\caption{\label{tab: BtoBcBcbar III}
$\overline B_q\to \B_c({\bf \bar 3_f})\overline \B_c ({\bf \bar 6_f})$ decay amplitudes in $\Delta S=-1$ and  $\Delta S=0$ transitions.}
\footnotesize{
\begin{ruledtabular}
\begin{tabular}{lclc}
Mode
          & $A \left(\overline B_q\to \B_c({\bf \bar 3_f})\overline \B_c ({\bf \bar 6_f})\right)$
          & Mode
          & $A \left(\overline B_q\to \B_c({\bf \bar 3_f})\overline \B_c ({\bf \bar 6_f})\right)$
          \\
\hline
          $B^-\to \Xi_c^+ \bar \Sigma_c^{--}$
          & $\sqrt2 C_{3,v}$ 
          & $B^-\to \Xi_c^0 \bar \Sigma_c^{-}$
          & $-C_{3,v}$ 
          \\     
          $\overline B{}^0\to \Xi_c^0 \bar \Sigma_c^{0}$
          & $-\sqrt2 C_{3,v}$
          & $\overline B{}^0\to \Xi_c^+ \bar \Sigma_c^{-}$
          & $C_{3,v}$
          \\
          $\overline B{}_s^0\to \Xi_c^0 \bar \Xi_c^{\prime 0}$
          & $-C_{3,vs}$
          & $\overline B{}_s^0\to \Xi_c^+ \bar \Xi_c^{\prime -}$
          & $C_{3,vs}$
          \\
\hline
         $B^-\to \Lambda_c^+ \bar \Sigma_c^{- -}$
          & $-\sqrt2 C'_3$
          & $B^-\to \Xi_c^0 \bar \Xi_c^{\prime -}$
          & $C'_{3,c}$
          \\
          $\overline B{}^0\to \Lambda_c^+ \bar \Sigma_c^{-}$
          & $-C'_3$
          & $\overline B{}^0\to \Xi_c^0 \bar \Xi_c^{\prime 0}$
          & $C'_{3,c}$ 
          \\          
          $\overline B{}^0_s\to \Lambda_c^+ \bar \Xi_c^{\prime -}$
          & $-C'_{3,s}$         
          & $\overline B{}_s^0\to \Xi_c^0 \bar \Omega_c^{0}$
          & $\sqrt2 C'_{3,cs}$
          \\                                            
\end{tabular}
\end{ruledtabular}
}
\end{table}

\begin{table}[t!]
\caption{\label{tab: BtoBcBcbar IV}
$\overline B_q\to \B_c({\bf 6_f})\overline \B_c ({\bf \bar 6_f})$ decay amplitudes in $\Delta S=-1$ and  $\Delta S=0$ transitions.}
\footnotesize{
\begin{ruledtabular}
\begin{tabular}{lclc}
Mode
          & $A \left(\overline B_q\to \B_c({\bf 6_f})\overline \B_c ({\bf \bar 6_f})\right)$
          & Mode
          & $A \left(\overline B_q\to \B_c({\bf 6_f})\overline \B_c ({\bf \bar 6_f})\right)$
          \\
\hline
           $B^-\to \Xi_c^{\prime +} \bar \Sigma_c^{--}$
          & $\sqrt2 C_{4,v}$ 
          & $B^-\to \Xi_c^{\prime 0} \bar \Sigma_c^{-}$
          & $C_{4,v}$ 
          \\     
          $B^-\to \Omega_c^{0} \bar \Xi_c^{-}$
          & $\sqrt2 C_{4,vc}$ 
          & $\overline B{}^0\to \Xi_c^{\prime +} \bar \Sigma_c^{-}$
          & $C_{4,v}$
          \\
          $\overline B{}^0\to \Xi_c^{\prime 0} \bar \Sigma_c^{0}$
          & $\sqrt2 C_{4,v}$
          & $\overline B{}^0\to \Omega_c^{0} \bar \Xi_c^{0}$
          & $\sqrt2 C_{4,vc}$
          \\
          $\overline B{}_s^0\to \Sigma_c^{++} \bar \Sigma_c^{--}$
          & $E_{4,v}$
          & $\overline B{}_s^0\to \Sigma_c^{+} \bar \Sigma_c^{-}$
          & $E_{4,v}$
          \\
         $\overline B{}_s^0\to \Sigma_c^{0} \bar \Sigma_c^{0}$
          & $E_{4,v}$
          & $\overline B{}_s^0\to \Xi_c^{\prime +} \bar \Xi_c^{\prime -}$
          & $C_{4,vs}+E_{4,vc}$
          \\
          $\overline B{}_s^0\to \Xi_c^{\prime 0} \bar \Xi_c^{\prime 0}$
          & $C_{4,vs}+E_{4,vc}$
          & $\overline B{}_s^0\to \Omega_c^{0} \bar \Omega_c^{0}$
          & $2C_{4,vcs}+E_{4,vcc}$
          \\
\hline
         $B^-\to \Sigma_c^+ \bar \Sigma_c^{- -}$
          & $\sqrt2 C'_4$
          & $B^-\to \Sigma_c^0 \bar \Sigma_c^{ -}$
          & $\sqrt2 C'_4$
          \\
          $B^-\to \Xi_c^{\prime 0} \bar \Xi_c^{\prime -}$
          & $C'_{4,c}$
          & $\overline B{}^0\to \Sigma_c^{++} \bar \Sigma_c^{ --}$
          & $E'_4$
          \\
          $\overline B{}^0\to \Sigma_c^+ \bar \Sigma_c^{-}$
          & $C'_4+E'_4$
          & $\overline B{}^0\to \Sigma_c^0 \bar \Sigma_c^{0}$
          & $2 C'_4+E'_4$
          \\
          $\overline B{}^0\to \Xi_c^{\prime +} \bar \Xi_c^{\prime -}$
          & $E'_{4,c}$ 
          & $\overline B{}^0\to \Xi_c^{\prime 0} \bar \Xi_c^{\prime 0}$
          & $C'_{4,c}+E'_{4,c}$ 
          \\          
          $\overline B{}^0\to \Omega_c^0 \bar \Omega_c^{0}$
          & $E'_{4,cc}$  
          & $\overline B{}_s^0\to \Sigma_c^+ \bar \Xi_c^{-}$
          & $C'_{4,s}$
          \\
          $\overline B{}_s^0\to \Sigma_c^0 \bar \Xi_c^{0}$
          & $\sqrt2 C'_{4,s}$
          & $\overline B{}_s^0\to \Xi_c^{\prime 0} \bar \Omega_c^{0}$
          & $\sqrt2 C'_{4,cs}$
          \\                                              
\end{tabular}
\end{ruledtabular}
}
\end{table}

\subsection{Branching ratios}

In the following numerical study, we will encounter $\overline B_q\to \B_c(J=\frac{1}{2}) \bar\B_c(J=\frac{1}{2})$ 
and $\overline B_q\to \B_c(J=\frac{3}{2}) \bar\B_c(J=\frac{1}{2})$ decays,
where $\bar\B_c(J=\frac{1}{2})$ is a spin-1/2 anti-charmed baryon, while $\B_c(J=\frac{1}{2}, \frac{3}{2})$ are spin-1/2 and spin-3/2 charmed baryons, respectively. 

For a $\overline B_q\to \B_c(1/2) \bar\B_c(1/2)$ decay, we can use the following formula for the branching ratio, see Appendix~\ref{App: Br},
\be
Br\left(\overline B_q\to \B_c\left(\frac{1}{2}\right)\bar\B_c\left(\frac{1}{2}\right)\right)
=\tau_{B_q} 
\frac{p_{cm}}{8 \pi m_{B_q}^2} 
\left|A\left(\overline B_q\to \B_c\left(\frac{1}{2}\right)\bar\B_c\left(\frac{1}{2}\right)\right)\right|^2,
\label{eq: rate 1/2 1/2}
\en
where $|A\left(\overline B_q\to \B_c\left(\frac{1}{2}\right)\bar\B_c\left(\frac{1}{2}\right)\right)|^2$ should contain various factors.
For example, for $B^-\to \Xi_c^0 \bar \Lambda_c^{-}$ and $B^-\to \Xi_c^0 \bar \Xi_c^{-}$ decays, we have (see also Table~\ref{tab: BtoBcBcbar I}),
\be
|A(B^-\to \Xi_c^0 \bar \Lambda_c^{-})|^2&=&|C_{1,v}|^2
=  \left|\frac{G_F}{\sqrt2} m_{B_q} V_{cb} V^*_{cs}\right|^2 c^2_{1,v},
\non\\
|A(B^-\to \Xi_c^0 \bar \Xi_c^{-})|^2&=&|C'_{1,c}|^2
=\left|\frac{G_F}{\sqrt2} m_{B_q} V_{cb} V^*_{cd}\right|^2  c^2_{1,c},
\en
with $c_{1,v}=(1+\delta^{c_1}_v) c_{1}$ and $c_{1,c}=(1+\delta^{c_1}_c) c_{1}$. 

Note that the above amplitudes squared are indeed the polarization sum of $|\bar u (a+\gamma_5 b) v|^2$,
see Eq. (\ref{eq: A1}).
Indeed, we have 
\be
 \left|\frac{G_F}{\sqrt2}  V_{cb} V^*_{cs}\right|^2 m^2_{B_q} c^2_{1,v}&=&
(2 m_{B_q}^2 - 2 (m_{\B_c}+ m_{\bar \B_c})^2) |a|^2+
(2 m_{B_q}^2 - 2 (m_{\B_c}- m_{\bar\B_c})^2) |b|^2
\non\\
&=&
(2 m_{B_q}^2 - 2 (m_{\B_c}- m_{\bar\B_c})^2) 
\left(|b|^2
+\frac{4 p_{cm}^2 m_{B_q}^2}
{[m_{B_q}^2 -  (m_{\B_c}- m_{\bar \B_c})^2]^2}
 |a|^2\right).
 \non\\
\en
The $|b|^2$ term is the $s$-wave term, while the $|a|^2$ term is the $p$-wave term.
The latter contribution is relatively kinematically suppressed in a typical
$\overline B_q\to \B_c(\frac{1}{2}) \bar\B_c(\frac{1}{2})$ decay. 
Consequently, the $s$-wave or the $|b|^2$ term is dominating.
To determine $s$-wave and $p$-wave contributions, 
one needs, for example, the asymmetry $\alpha$.
Since only data on total rates are available presently,
we use, for simplicity, a single parameter, namely $c^2_{1,v}$, for the amplitudes squared in this study.
This situation can be improved when data on asymmetries becomes available.  

Furthermore, although the factor $(m_{B_q}^2 - (m_{\B_c}- m_{\bar\B_c})^2)$ in the above equation
can introduce some SU(3) breaking effects, in the modes studied in this work, the SU(3) breaking effects are at most at 1\% level.
As we shall see, this is much less than the SU(3) breaking effects in amplitudes introduced earlier. 
Hence, it is not necessary to modify the $m_{B_q}^2$ factor before $c^2_{1,v}$.

For a $\overline B_q\to \B_c(\frac{3}{2}) \bar\B_c(\frac{1}{2})$ decay, we can use the following formula for the branching ratio, see Appendix~\ref{App: Br},
\be
Br\left(\overline B_q\to \B_c\left(\frac{3}{2}\right)\bar\B_c\left(\frac{1}{2}\right)\right)
=\tau_{B_q} 
\frac{p_{cm}}{8 \pi m_{B_q}^2} \left(\frac{p_{cm}}{m_{\B_c(3/2)}} \right)^2
\left|A\left(\overline B_q\to \B_c\left(\frac{3}{2}\right)\bar\B_c\left(\frac{1}{2}\right)\right)\right|^2.
\label{eq: rate 3/2 1/2}
\en
Note that we show explicitly the additional $(p_{cm}/m_{\B_c(3/2)})^2$ factor in the above equation,
as the decay amplitude for a $\overline B_q\to \B_c(\frac{3}{2}) \bar\B_c(\frac{1}{2})$ decay is basically, $(p_{cm}/m_{\B_c(3/2)}) \bar u_1(a+\gamma_5 b) v_2$,
see Eq. (\ref{eq: 3/2 1/2}).
Similarly, the amplitude squared, $|A\left(\overline B_q\to \B_c\left(\frac{3}{2}\right)\bar\B_c\left(\frac{1}{2}\right)\right)|^2$, should contain various factors.
For example, for $B^-\to \Xi_c^{0}(2645) \bar \Lambda_c^{-}$ and $\overline B{}^0\to \Sigma_c^+ (2520)\bar \Lambda_c^{-}$ decays, similarly, we have
\be
|A(B^-\to \Xi_c^{0}(2645) \bar \Lambda_c^{-})|^2
&=&|-\tilde C_{2,v}|^2
=  \left|\frac{G_F}{\sqrt2} m_{B_q} V_{cb} V^*_{cs}\right|^2 \tilde c^2_{2,v},
\non\\
|A(\overline B{}^0\to \Sigma_c^+ (2520)\bar \Lambda_c^{-})|^2
&=&|\tilde C'_{2}|^2
=\left|\frac{G_F}{\sqrt2} m_{B_q} V_{cb} V^*_{cd}\right|^2  \tilde c^2_{2},
\en
with $\tilde c_{2,v}=(1+\delta^{\tilde c_2}_v) \tilde c_{2}$.
Note that $\Xi_c^{0}(2645)$ and $\Sigma_c^+ (2520)$ are spin-3/2 charmed baryons in a ${\bf 6_f}$ multiple, see Table~\ref{tab: quantum number}.
Hence, the decompositions of their amplitudes are similar to those given in Table~\ref{tab: BtoBcBcbar II}.

Note that the $p_{cm}$ and $p_{cm}^3$ kinematic factors in Eqs. (\ref{eq: rate 1/2 1/2}) and (\ref{eq: rate 3/2 1/2}), respectively, provide additional SU(3) breaking effects in rates.
These effects can be sizable in some modes.

\section{Numerical results}

Numerical results on rates of $\overline B_q\to \B_c({\bf \bar 3_f})\overline \B_c ({\bf 3_f})$,  
$\overline B_q\to \B_c({\bf 6_f})\overline \B_c ({\bf 3_f})$
and 
$\overline B_q\to \B_c({\bf \bar 3_f})\overline \B_c ({\bf \bar 6_f})$ decays
will be presented in this section.
As one can see from Tables~\ref{tab: expt} and \ref{tab: quantum number},
data on the rates of some of these modes are available, 
while no data on $\overline B_q\to \B_c({\bf 6_f})\overline \B_c ({\bf \bar 6_f})$ decays is reported yet.
Hence, we do not include the latter modes in this numerical study.
Masses and lifetimes of $B_q$ mesons and masses of $\B_c$ baryons are taken from ref. \cite{ParticleDataGroup:2024cfk}.
CKM matrix elements are from the latest fit in ref. \cite{CKMfitter}.

\subsection{$\overline B_q\to \B_c({\bf \bar 3_f})\overline \B_c ({\bf 3_f})$ decay rates}

In this subsection, we give the numerical results of the branching ratios of $\overline B_q\to \B_c({\bf \bar 3_f})\overline \B_c ({\bf 3_f})$ decays for low-lying $\B_c({\bf \bar 3_f})$ and $\overline \B_c ({\bf 3_f})$, and for excited $\B_c({\bf \bar 3_f})$ with $J^P=\frac{1}{2}^-$, namely $\overline B_q\to\Xi_c^{0,+}(2790)\bar\B_c ({\bf 3_f})$ and $\Lambda_c^+(2595)\bar\B_c ({\bf 3_f})$ decays. 

\subsubsection{Low-lying case}

We start with the low-lying case.
Presently, we have data on rates of the following four modes, namely
$B^-\to \Xi_c^0 \bar \Lambda_c^{-}$,
$\overline B{}^0\to \Xi_c^+ \bar \Lambda_c^{-}$,
$\overline B{}_s^0\to \Lambda_c^+ \bar \Lambda_c^{-}$
and $\overline B{}^0\to \Lambda_c^+ \bar \Lambda_c^{-}$ decays with rates shown in Table~\ref{tab: expt}.
Their amplitudes are governed by 
$C_{1,v}$,
$E_{1,v}$
and
$C'_1+E'_1$,
see Table~\ref{tab: BtoBcBcbar I},
which can be expressed in terms of four unknown parameters,
$c_1$, 
$e_1$,
$\delta_v^{c_1}$,
and 
$\delta_v^{e_1}$.
At first sight, it seems that these four unknown parameters can be fully determined from the above four experimental inputs.
This is, however, not the case, as $B^-\to \Xi_c^0 \bar \Lambda_c^{-}$ and $\overline B{}^0\to \Xi_c^+ \bar \Lambda_c^{-}$ decays are
isospin-related, which are identical in the isospin limit, and can hardly be counted as two independent inputs.   
To proceed, one needs to impose a constraint on these parameters. 
Since $\delta_v^{c_1}$ and $\delta_v^{e_1}$ are measurements of the SU(3) breaking effects, they are expected to be of similar size.
For simplicity, we assume
\be
|\delta_v^{c_1}|=|\delta_v^{e_1}|.
\label{eq: constraint}
\en 
Note that this working assumption can be relaxed when data from more modes becomes available. 

The above constraint can be reduced to the two following cases,
\be
&&{\rm case\, 1}:
\quad
\delta_v^{c_1}=+\delta_v^{e_1},
\label{eq: case 1}
\non\\
&&{\rm case\, 2}:
\quad
\delta_v^{c_1}=-\delta_v^{e_1}.
\label{eq: case 2}
\en
From $\chi^2$-fit using rates of the above four modes, we obtain for the best fits,
\be
&&{\rm case\, 1}:
\quad
c_1=0.15, 
\,
e_1=-0.03,
\,
\delta_v^{c_1}=+\delta_v^{e_1}=0.88, 
\non\\
&&{\rm case\, 2}:
\quad
c_1=0.21, 
\,
e_1=-0.09,
\,
\delta_v^{c_1}=-\delta_v^{e_1}=0.35,
\en
both with $\chi^2_{\rm min}=0.315$.

From these results, one can immediately see that case 1 is unfavorable, as the fit indicates a significant SU(3) breaking effect, with 88\% SU(3) breaking in amplitudes.
On the contrary, the fit in case 2 needs only 35\% SU(3) breaking effects. It is more reasonable and acceptable compared to case 1. 
Consequently, we will proceed with case 2 in the following study of the low-lying $\overline B_q\to \B_c({\bf \bar 3_f})\overline \B_c ({\bf 3_f})$ decays.

\begin{table}[t!]
\caption{\label{tab: c1}
Parameters in $\overline B_q\to \B_c({\bf \bar 3_f})\overline \B_c ({\bf 3_f})$ for low lying $\B_c({\bf \bar 3_f})$ and $\overline \B_c ({\bf 3_f})$, 
where $c_1$,  $e_1$, $\delta^{c_1}_v$ are fitted parameters with a constraint on SU(3) breaking parameters, $\delta^{e_1}_v=-\delta^{c_1}_v$. 
Their uncertainties are obtained by scanning parameter space with $\chi^2\leq \chi^2_{\rm min.}+1$.
We assume other SU(3) breaking parameters ($\delta^{c_1}_c$, $\delta^{c_1}_s$, $\delta^{e_1}_c$) have sizes at most as $|\delta^{c_1}_v|$ and $|\delta^{e_1}_v|$.}
\begin{ruledtabular}
\begin{tabular}{cccc} 
 $c_1$
          & $e_1$
          & $\delta^{c_1}_v$ 
          & $\delta^{e_1}_v(=-\delta^{c_1}_v)$
          \\
\hline 
$0.209_{-0.016}^{+0.015}$
         & $-0.091\pm 0.015$
         & $0.35\pm 0.11$
         & $-0.35\pm 0.11$ 
          \\          
\hline 
$\delta^{c_1}_c$
          & $\delta^{c_1}_s$
          & $\delta^{e_1}_c$
          & $\chi^2_{\rm min.}$
         \\ \hline
$0\pm 0.35$
          & $0\pm 0.35$
          & $0\pm 0.35$
          & $0.315$
          \\
\end{tabular}
\end{ruledtabular}
\end{table}

We show in Table \ref{tab: c1}, the fitted values of $c_1$, $e_1$, $\delta_v^{c_1}$ and $\delta_v^{e_1}$ with uncertainties obtained 
by scanning the parameter space allowed by $\chi^2\leq \chi^2_{\rm min.}+1$.
As noted, the four experimental rates are input of the fit, and they are insensitive to the other three SU(3) breaking parameters, namely
$\delta^{c_1}_c$, $\delta^{c_1}_s$ and $\delta^{e_1}_c$. 
These $\delta$s likely have sizes similar to $\delta_v^{c_1}$ and $\delta_v^{e_1}$.
We assume their values as $0\pm 0.35$ independently.

There are three implications from the fit.
First, the $W$-exchange diagram is sizable in $\overline B_q\to \B_c({\bf \bar 3_f})\overline \B_c ({\bf 3_f})$ decay, 
with $|E^{(\prime)}_1|/|C^{(\prime)}_1|$ ratio about 44\%.
Neglecting the contribution from the exchange-$W$ tree diagram is not acceptable.
Second, there is a large cancellation between the internal $W$-tree and the exchange-$W$-tree amplitudes.   
Third, SU(3) breaking is sizable. 
We find that 35\% SU(3) breaking in amplitudes is needed.
Furthermore, they work differently in different amplitudes.
With the SU(3) breaking effects,
the internal $W$-tree amplitude $C_{1,v}$ is enlarged, 
while the exchange-$W$ tree amplitude $E_{1,v}$ is reduced.
It remains to be seen if this pattern is followed by other internal $W$-tree and exchange-$W$ tree amplitudes,
such as $C_{1,s}$, $C_{1,c}$, $C_{1,vs}$, $E_{1,s}$ and $E_{1,vc}$. 

Note that some of the above observations on the low-lying case have been pointed out in ref.~\cite{LHCb:2025ueu}, but in this work, more detailed information, such as the size of the ratio of the topological amplitudes, the sizes of the SU(3) breaking effects, and different behaviors of the effects on different topological amplitudes, is provided.

Note that in principle the phases of $E_1$ and $C_1$ need not be the same. However, from the large cancellation of $C_1$ and $E_1$, 
and the constraint of the size of $E_1$ from the $W$-exchange mode, the data support the case that they have similar phase or more precisely, opposite phases.
In other words, adding the imaginary part of $E_1$ does not improve the fit. 

\begin{table}[t!]
\caption{\label{tab: Br BtoBcBcbar 3bar3}
$\overline B_q\to \B_c({\bf \bar 3_f})\overline \B_c ({\bf 3_f})$ decay amplitudes and branching ratios (in unit of $10^{-4}$) in $\Delta S=-1$ and  $\Delta S=0$ transitions. Experimental results are inputs of the $\chi^2$-fit. 
Results from ref.~\cite{Hsiao:2023mud} are also shown for comparison.
}
\footnotesize{
\begin{ruledtabular}
\begin{tabular}{lcccc}
Mode
          & $A \left(\overline B_q\to \B_c({\bf \bar 3_f})\overline \B_c ({\bf 3_f})\right)$
          & $Br \left(\overline B_q\to \B_c({\bf \bar 3_f})\overline \B_c ({\bf 3_f})\right)$
          & $Br \left(\overline B_q\to \B_c({\bf \bar 3_f})\overline \B_c ({\bf 3_f})\right)$
          & Expt.
           \\
          & 
          & this work
          & ref. \cite{Hsiao:2023mud}
          & 
           \\
\hline
$B^-\to \Xi_c^0 \bar \Lambda_c^{-}$
          & $C_{1,v}$  
          & $10.06_{-2.04}^{+1.98}$
          & $7.8^{+2.3}_{-2.0}$
          & $9.51\pm 2.28$~\cite{Belle:2018kzz}
          \\
$\overline B{}^0\to \Xi_c^+ \bar \Lambda_c^{-}$
          & $C_{1,v}$
          & $9.34_{-1.90}^{+1.84}$
          & $7.2^{+2.1}_{-1.9}$
          & $11.6\pm 4.46$~\cite{Belle:2019bgi}
          \\
$\overline B{}_s^0\to \Lambda_c^+ \bar \Lambda_c^{-}$
          & $-E_{1,v}$
          & $0.50_{-0.15}^{+0.16}$
          & $0.81^{+0.17}_{-0.15}$
          & $0.50\pm 0.16$~ \cite{LHCb:2025ueu}
          \\
$\overline B{}_s^0\to \Xi_c^0 \bar \Xi_c^{0}$
          & $-C_{1,vs}-E_{1,vc}$
          & $5.33_{-4.44}^{+8.67}$
          & $3.0^{+1.4}_{-1.1}$
          & $-$
          \\     
$\overline B{}_s^0\to \Xi_c^+ \bar \Xi_c^{-}$
          & $-C_{1,vs}-E_{1,vc}$
          & $5.36_{-4.46}^{+8.72}$
          & $3.0^{+1.4}_{-1.1}$
          & $-$
          \\
           \hline
$\overline B{}^0\to \Lambda_c^+ \bar \Lambda_c^{-}$
          & $-C'_1-E'_1$
          & $0.101_{-0.032}^{+0.031}$
          & $0.21^{+0.10}_{-0.08}$
          & $0.101\pm0.032$
          \\          
$\overline B{}^0\to \Xi_c^0 \bar \Xi_c^{0}$
          & $-C'_{1,c}-E'_{1,c}$  
          & $0.071_{-0.071}^{+0.227}$
          & $0.15^{+0.07}_{-0.06}$
          & $-$
          \\
$B^-\to \Xi_c^0 \bar \Xi_c^{-}$
          & $C'_{1,c}$
          & $0.240_{-0.154}^{+0.261}$
          & $0.34^{+0.10}_{-0.09}$
          & $-$
          \\
$\overline B{}_s^0\to \Lambda_c^+ \bar \Xi_c^{-}$
          & $C'_{1,s}$
          & $0.297_{-0.190}^{+0.322}$
          & $0.39^{+0.12}_{-0.10}$
          & $-$
          \\                                            
$\overline B{}^0\to \Xi_c^+ \bar \Xi_c^{-}$
          & $-E'_{1,c}$  
          & $0.042_{-0.030}^{+0.063}$ 
          & $0.030\pm 0.006$       
          & $-$
          \\ 
\end{tabular}
\end{ruledtabular}
}
\end{table}

We show in Table \ref{tab: Br BtoBcBcbar 3bar3},
the branching ratios of
$\overline B_q\to \B_c({\bf \bar 3_f})\overline \B_c ({\bf 3_f})$ decays in $\Delta S=-1$ and  $\Delta S=0$ transitions. 
We also show the corresponding decay amplitudes taken from Table~\ref{tab: BtoBcBcbar I} for our convenience. 
The results in the table are obtained using the parameters in Table~\ref{tab: c1}. 
Experimental results are inputs of the $\chi^2$-fit.
Results from ref.~\cite{Hsiao:2023mud} are also shown for comparison.

From the table, we see that the central value of the fitted $B^-\to \Xi_c^0 \bar \Lambda_c^{-}$ rate is slightly higher than the central value of the experimental result,
while the case of the $\overline B{}^0\to \Xi_c^+ \bar \Lambda_c^{-}$ decay is the other way around. 
The amplitudes of these two modes are both $C_{1,v}$.
Hence, the maximum of $|C_{1,v}|$ is constrained by the $B^-\to \Xi_c^0 \bar \Lambda_c^{-}$ data, 
while the minimum is by the $\overline B{}^0\to \Xi_c^+ \bar \Lambda_c^{-}$ data.
For $\overline B{}_s^0\to \Lambda_c^+ \bar \Lambda_c^{-}$ and $\overline B{}^0\to \Lambda_c^+ \bar \Lambda_c^{-}$ decays, the fit reproduces the data well.

We have predictions on six other modes, namely 
$\overline B{}_s^0\to \Xi_c^0 \bar \Xi_c^{0}$,
$\overline B{}_s^0\to \Xi_c^+ \bar \Xi_c^{-}$,
$\overline B{}^0\to \Xi_c^0 \bar \Xi_c^{0}$,
$B^-\to \Xi_c^0 \bar \Xi_c^{-}$,
$\overline B{}_s^0\to \Lambda_c^+ \bar \Xi_c^{-}$
and
$\overline B{}^0\to \Xi_c^+ \bar \Xi_c^{-}$
decays.
However, the uncertainties in these predicted rates are substantial.
The reason can be traced to the poorly known SU(3) breaking parameters, $\delta^{c_1}_c$, $\delta^{c_1}_s$, $\delta^{e_1}_c$, which are not constrained by the present data.
From the amplitudes shown in the same table, these six modes indeed depend on these parameters.

The uncertainties in the rates of these modes simply reflect our estimates of the SU(3)- breaking parameters. 
As stated previously, their sizes are estimated to be at most as $|\delta^{c_1}_v|$ and $|\delta^{e_1}_v|$.
This observation, in fact, identifies the opportunity to really constrain or even determine these SU(3) breaking parameters by measuring these modes,
as they are highly sensitive to them.

For example, in $B^-\to \Xi_c^0 \bar \Xi_c^{-}$ decay, we have $A(B^-\to \Xi_c^0 \bar \Xi_c^{-})=C'_{1,c}=C'_1 (1+\delta^{c_1}_c)$.
Although the present data can effectively constrain $C'_1$, see the $c_1$ value in Table~\ref{tab: c1}, the value of $\delta^{c_1}_c$ is basically unconstrained.
The predicted $B^-\to \Xi_c^0 \bar \Xi_c^{-}$ rate has large uncertainty, 
giving $Br(B^-\to \Xi_c^0 \bar \Xi_c^{-})=(0.240_{-0.154}^{+0.261})\times 10^{-4}$, 
reflecting our poor understanding on the SU(3) breaking parameter $\delta^{c_1}_c$.
Therefore, a measured $B^-\to \Xi_c^0 \bar \Xi_c^{-}$ rate can give valuable information on $\delta^{c_1}_c$.

The comparison of our results to those obtained in ref.~\cite{Hsiao:2023mud}, as shown in Table \ref{tab: Br BtoBcBcbar 3bar3}, makes the above point even clearer.
The uncertainties of our predicted results on
$\overline B{}_s^0\to \Xi_c^0 \bar \Xi_c^{0}$,
$\overline B{}_s^0\to \Xi_c^+ \bar \Xi_c^{-}$,
$\overline B{}^0\to \Xi_c^0 \bar \Xi_c^{0}$,
$B^-\to \Xi_c^0 \bar \Xi_c^{-}$,
$\overline B{}_s^0\to \Lambda_c^+ \bar \Xi_c^{-}$
and
$\overline B{}^0\to \Xi_c^+ \bar \Xi_c^{-}$
decay rates
are almost always larger than those obtained in ref.~\cite{Hsiao:2023mud}.
This discrepancy clearly shows the different treatments of SU(3) breaking effects in amplitudes, which are absent in ref.~\cite{Hsiao:2023mud}.
Hence, measuring these rates can constrain or even identify the underlying SU(3) breaking effects.

\subsubsection{Excited $\B_c({\bf \bar 3_f})$ case}

\begin{table}[t!]
\caption{\label{tab: c1p}
Parameters in $\overline B_q\to \B_c({\bf \bar 3_f})\overline \B_c ({\bf 3_f})$ for $1/2^-$ $\B_c({\bf \bar 3_f})$ and low-lying $\overline \B_c ({\bf 3_f})$, 
where $\tilde c_{1,v}\equiv (1+\delta^{\tilde c_1}_v)\tilde c_1$ is the fitted parameter. 
The uncertainty is obtained by scanning $\chi^2\leq \chi^2_{\rm min.}+1$.
We assume SU(3) breaking parameters $\delta^{\tilde c_1}_v$, $\delta^{\tilde c_1}_c$ and $\delta^{\tilde c_1}_s$ have sizes at most as $|\delta^{c_1}_v|$.
Information about $\tilde e_1$ and its SU(3) breaking parameters is unavailable from the present data.}
\begin{ruledtabular}
\begin{tabular}{cccc} 
 $\tilde c_{1,v}\equiv (1+\delta^{\tilde c_1}_v)\tilde c_1$
          & $\delta^{\tilde c_1}_v$
          & $\delta^{\tilde c_1}_c$ 
          & $\delta^{\tilde c_1}_s$
          \\
\hline 
$0.118_{-0.024}^{+0.020}$
         & $0\pm 0.35$
         & $0\pm 0.35$
         & $0\pm 0.35$
          \\          
\hline 
$\tilde e_1$
          & $\delta^{\tilde e_1}_v$
          & $\delta^{\tilde e_1}_c$
          & $\chi^2_{\rm min.}$
         \\ \hline
$-$
          & $-$
          & $-$
          & $0$
          \\
\end{tabular}
\end{ruledtabular}
\end{table}

We now give the numerical results of the branching ratios of $\overline B_q\to\Xi_c^{0,+}(2790)\bar\B_c ({\bf 3_f})$ and $\Lambda_c^+(2595)\bar\B_c ({\bf 3_f})$ decays with $\bar\B_c ({\bf 3_f})$ a low-lying anti-charmed state. 
As noted in Table~\ref{tab: quantum number}, $\Lambda_c^+(2595)$ and $\Xi_c^{0,+}(2790)$ are $J^P=\frac{1}{2}^-$ charmed baryons which form a ${\bf \bar 3_f}$ multiplet.
Hence, these decays are also $\overline B_q\to \B_c({\bf \bar 3_f})\overline \B_c ({\bf 3_f})$ decays, and their decay amplitudes can be decomposed similarly as those shown in Table \ref{tab: BtoBcBcbar I}.

\begin{table}[t!]
\caption{\label{tab: Br BtoBcBcbar 3barp3}
$\overline B_q\to \B_c({\bf \bar 3_f})\overline \B_c ({\bf 3_f})$ decay amplitudes and branching ratios (in unit of $10^{-4}$) in $\Delta S=-1$ and  $\Delta S=0$ transitions,
where $\B_c({\bf \bar 3_f})$ are $1/2^-$ states, while $\overline \B_c ({\bf 3_f})$ are low-lying states.
The experimental result is the input of the $\chi^2$-fit. As the information about $\tilde e_1$ is unavailable, there is no prediction on rates of modes involving $\tilde E^{(\prime)}_1$.}
\footnotesize{
\begin{ruledtabular}
\begin{tabular}{lccc}
Mode
          & $A \left(\overline B_q\to \B_c({\bf \bar 3_f})\overline \B_c ({\bf 3_f})\right)$
          & $Br \left(\overline B_q\to \B_c({\bf \bar 3_f})\overline \B_c ({\bf 3_f})\right)$
          & Expt.
           \\
\hline
$B^-\to \Xi_c^0(2790) \bar \Lambda_c^{-}$
          & $\tilde C_{1,v}$  
          & $1.10\pm 0.40$
          & $1.1\pm 0.4$ \cite{Belle:2019pze}
          \\
$\overline B{}^0\to \Xi_c^+(2790) \bar \Lambda_c^{-}$
          & $\tilde C_{1,v}$
          & $1.02\pm 0.37$
          & $-$
          \\
$\overline B{}_s^0\to \Lambda_c^+(2595) \bar \Lambda_c^{-}$
          & $-\tilde E_{1,v}$
          & $-$
          & $-$
          \\
$\overline B{}_s^0\to \Xi_c^0(2790) \bar \Xi_c^{0}$
          & $-\tilde C_{1,vs}-\tilde E_{1,vc}$
          & $-$
          & $-$
          \\
$\overline B{}_s^0\to \Xi_c^+(2790) \bar \Xi_c^{-}$
          & $-\tilde C_{1,vs}-\tilde E_{1,vc}$
          & $-$
          & $-$
          \\
\hline
$B^-\to \Xi_c^0(2790) \bar \Xi_c^{-}$
          & $\tilde C'_{1,c}$
          & $0.018_{-0.015}^{+0.087}$
          & $-$
          \\
$\overline B{}_s^0\to \Lambda_c^+(2595) \bar \Xi_c^{-}$
          & $\tilde C'_{1,s}$
          & $0.096_{-0.082}^{+0.471}$
          & $-$
          \\                  
$\overline B{}^0\to \Lambda_c^+(2595) \bar \Lambda_c^{-}$
          & $-\tilde C'_1-\tilde E'_1$
          & $-$
          & $-$
          \\          
$\overline B{}^0\to \Xi_c^0(2790) \bar \Xi_c^{0}$
          & $-\tilde C'_{1,c}-\tilde E'_{1,c}$  
          & $-$
          & $-$
          \\                          
$\overline B{}^0\to \Xi_c^+(2790) \bar \Xi_c^{-}$
          & $-\tilde E'_{1,c}$  
          & $-$         
          & $-$
          \\ 
\end{tabular}
\end{ruledtabular}
}
\end{table}

On the experimental side, presently only the $B^-\to\Xi_c^{0}(2790)\bar \Lambda_c^-$ rate is measured, see Table~\ref{tab: expt}.
Note that the amplitude of this mode is given by
\be
A(B^-\to \Xi_c^0(2790) \bar \Lambda_c^{-})=\tilde C_{1,v},  
\en
where the tilde on top of the topological amplitude is used to differentiate the internal $W$-tree amplitude in this case from that in the low-lying case. 
As this is the only mode measured, we can only have information on the internal $W$-tree amplitude with SU(3) breaking effect on the vertex, namely $\tilde C_{1,v}=(1+\delta^{\tilde c_1}_v) \tilde C_{1}$ or effectively $\tilde c_{1,v}=(1+\delta^{\tilde c_1}_v) \tilde c_{1}$.
The fitted value on $\tilde c_{1,v}$ using the measured $B^-\to\Xi_c^{0}(2790)\bar \Lambda_c^-$ rate is given in Table~\ref{tab: c1p}.  
The uncertainty is obtained by scanning $\chi^2\leq \chi^2_{\rm min.}+1$, and it simply reflects the uncertainty of the measured $B^-\to\Xi_c^{0}(2790)\bar \Lambda_c^-$ rate.
We assume that SU(3) breaking parameters $\delta^{\tilde c_1}_v$, $\delta^{\tilde c_1}_c$ and $\delta^{\tilde c_1}_s$ have sizes at most as $|\delta^{c_1}_v|$ in the low-lying case.
Information about $\tilde e_1$ and its SU(3) breaking parameters is unavailable from the present data.

We give predictions on the rates of some related $\overline B_q\to\Xi_c^{0,+}(2790)\bar\B_c ({\bf 3_f})$ and $\Lambda_c^+(2595)\bar\B_c ({\bf 3_f})$ modes in Table~\ref{tab: Br BtoBcBcbar 3barp3}.
The corresponding decay amplitudes are also shown for our convenience.
Since we do not have any information on the $W$-exchange amplitude in these modes, the decay rates of modes containing $\tilde E$ and $\tilde E'$ cannot be predicted. 
Only modes governed by internal $W$-tree amplitudes can be predicted, 
namely  
$\overline B{}^0\to \Xi_c^+(2790) \bar \Lambda_c^{-}$,
$B^-\to \Xi_c^0(2790) \bar \Xi_c^{-}$
and
$\overline B{}_s^0\to \Lambda_c^+(2595) \bar \Xi_c^{-}$ 
decays.
The $\overline B{}^0\to \Xi_c^+(2790) \bar \Lambda_c^{-}$ decay amplitude is also governed by $\tilde C_{1,v}$,
while the other two modes are governed by $\tilde C'_{1,c}$ and $\tilde C'_{1,s}$, respectively.
The $\overline B{}^0\to \Xi_c^+(2790) \bar \Lambda_c^{-}$ mode is the isospin related mode of $B^-\to \Xi_c^0(2790) \bar \Lambda_c^{-}$,
while the amplitudes of the other two modes can be expressed in terms of $\tilde C'_{1,v}$, which is related to $\tilde C_{1,v}$ through CKM factors and SU(3) breaking factors as follows, 
\be
\tilde C'_{1,c}=(1+\delta^{\tilde c_1}_c) \tilde C'_{1}=\frac{1+\delta^{\tilde c_1}_c}{1+\delta^{\tilde c_1}_v} \tilde C'_{1,v},
\quad
\tilde C'_{1,s}=(1+\delta^{\tilde c_1}_s) \tilde C'_{1}=\frac{1+\delta^{\tilde c_1}_s}{1+\delta^{\tilde c_1}_v} \tilde C'_{1,v}.
\en
With our estimations on these SU(3) breaking factors and the measured $B^-\to\Xi_c^{0}(2790)\bar \Lambda_c^-$ rate, the decay rates of these three decay modes are obtained accordingly and are shown in Table~\ref{tab: Br BtoBcBcbar 3barp3}.

\subsection{$\overline B_q\to \B_c({\bf 6_f})\overline \B_c ({\bf 3_f})$ decay rates}

We consider $\overline B_q\to \B_c({\bf 6_f})\overline \B_c ({\bf 3_f})$ decays for low-lying $\B_c({\bf 6_f})$ and $\overline \B_c ({\bf 3_f})$ and for excited $\B_c({\bf 6_f})$ in this subsection.

\subsubsection{Low-lying case}

\begin{table}[t!]
\caption{\label{tab: c2}
Parameters in $\overline B_q\to \B_c({\bf 6_f})\overline \B_c ({\bf 3_f})$ decays for low lying $\B_c({\bf 6_f})$ and $\overline \B_c ({\bf 3_f})$, 
where $c_{2,v}\equiv (1+\delta^{c_2}_v) c_2$ is a fitted parameter. 
The uncertainty is obtained by scanning $\chi^2\leq \chi^2_{\rm min.}+1$.
We assume SU(3) breaking parameters ($\delta^{c_2}_v$, $\delta^{c_2}_c$, $\delta^{c_2}_s$) have sizes at most as $|\delta^{c_1}_v|$.}
\begin{ruledtabular}
\begin{tabular}{ccccc} 
 $c_{2,v} \equiv (1+\delta^{c_2}_v) c_2$
          & $\delta^{c_2}_v$ 
          & $\delta^{c_2}_c$
          & $\delta^{c_2}_s$
          & $\chi^2_{\rm min.}$
         \\
          \hline
$0.173_{-0.062}^{+0.045}$
          & $0\pm 0.35$
          & $0\pm 0.35$
          & $0\pm 0.35$
          & $0$
          \\
\end{tabular}
\end{ruledtabular}
\end{table}

We start with the numerical results of $\overline B_q\to \B_c({\bf 6_f})\overline \B_c ({\bf 3_f})$ decays for low-lying $\B_c({\bf 6_f})$ and $\overline \B_c ({\bf 3_f})$ baryons.
Note that, as forbidden by charge conservation, we do not have the $\overline B_q\to \Sigma_c^{++} \overline \B_c ({\bf 3_f})$ decay.
The amplitudes are governed by internal $W$-tree amplitudes, which may contain SU(3) breaking effects, see Table~\ref{tab: BtoBcBcbar II}.
So far only the measurement of the $B^-\to \Xi_c^{\prime 0} \bar \Lambda_c^{-}$ rate among these modes is reported, see Table~\ref{tab: expt}. 
Its amplitude is given by $A(\overline B{}^0\to \Xi_c^{\prime +} \bar \Lambda_c^{-})=C_{2,v}$.
Hence, only $|C_{2,v}|$, or equivalently $|c_{2,v}|= |(1+\delta^{c_2}_v) c_2|$, can be constrained by data.
The fitted parameter is given in Table~\ref{tab: c2}.

\begin{table}[t!]
\caption{\label{tab: br BtoBcBcbar 63}
$\overline B_q\to \B_c({\bf 6_f})\overline \B_c ({\bf 3_f})$ decay amplitudes and branching ratios (in unit of $10^{-4}$) in $\Delta S=-1$ and  $\Delta S=0$ transitions.
The experimental result is the input of the $\chi^2$-fit.
}
\footnotesize{
\begin{ruledtabular}
\begin{tabular}{lccc}
Mode
          & $A \left(\overline B_q\to \B_c({\bf 6_f})\overline \B_c ({\bf 3_f})\right)$
          & $Br \left(\overline B_q\to \B_c({\bf 6_f})\overline \B_c ({\bf 3_f})\right)$
          & Expt
          \\
\hline
           $B^-\to \Xi_c^{\prime 0} \bar \Lambda_c^{-}$
          & $-C_{2,v}$
          & $3.40\pm 2.00$
          & $3.4\pm 2.0$ \cite{Belle:2019pze} 
          \\
$\overline B{}^0\to \Xi_c^{\prime +} \bar \Lambda_c^{-}$
          & $C_{2,v}$
          & $3.15\pm 1.85$
          & $-$ 
          \\
$B^-\to \Omega_c^{0} \bar \Xi_c^{-}$
          & $\sqrt2 C_{2,vc}$
          & $3.66_{-\,\,\,3.34}^{+10.10}$
          & $-$ 
          \\    
$\overline B{}^0\to \Omega_c^{0} \bar \Xi_c^{0}$
          & $-\sqrt2 C_{2,vc}$
          & $3.35_{-3.06}^{+9.25}$
          & $-$ 
          \\
$\overline B{}_s^0\to \Xi_c^{\prime +} \bar \Xi_c^{-}$
          & $-C_{2,vs}$
          & $2.81_{-2.56}^{+7.75}$
          & $-$ 
          \\
$\overline B{}_s^0\to \Xi_c^{\prime 0} \bar \Xi_c^{0}$
          & $C_{2,vs}$
          & $2.80_{-2.55}^{+7.70}$
          & $-$ 
           \\
\hline
         $B^-\to \Sigma_c^0 \bar \Lambda_c^{-}$
          & $-\sqrt2 C'_2$
          & $0.41_{-0.32}^{+1.14}$
          & $-$ 
          \\
$\overline B{}^0\to \Sigma_c^+ \bar \Lambda_c^{-}$
          & $C'_2$
          & $0.19_{-0.15}^{+0.53}$
          & $-$ 
          \\        
$B^-\to \Xi_c^{\prime 0} \bar \Xi_c^{-}$
          & $C'_{2,c}$
          & $0.14_{-0.12}^{+0.80}$
          & $-$ 
         \\
$\overline B{}^0\to \Xi_c^{\prime 0} \bar \Xi_c^{0}$
          & $-C'_{2,c}$ 
          & $0.13_{-0.11}^{+0.74}$
          & $-$ 
          \\
$\overline B{}_s^0\to \Sigma_c^+ \bar \Xi_c^{-}$
          & $-C'_{2,s}$
          & $0.18_{-0.16}^{+1.03}$
          & $-$ 
          \\
$\overline B{}_s^0\to \Sigma_c^0 \bar \Xi_c^{0}$
          & $\sqrt2 C'_{2,s}$
          & $0.35_{-0.32}^{+2.05}$
          & $-$ 
          \\                                            
\end{tabular}
\end{ruledtabular}
}
\end{table}

By assuming SU(3) breaking parameters ($\delta^{c_2}_v$, $\delta^{c_2}_c$, $\delta^{c_2}_s$) have sizes at most as $|\delta^{c_1}_v|$,
results of other $\overline B_q\to \B_c({\bf 6_f})\overline \B_c ({\bf 3_f})$ decay rates can be obtained and are shown in Table~\ref{tab: br BtoBcBcbar 63}.
Their amplitudes are also shown for our convenience.
Note that $\overline B{}^0\to \Xi_c^{\prime +} \bar \Lambda_c^{-}$ decay is the isospin-related mode to the measured mode, 
the uncertainty in rates originated from the uncertainty of the measured $B^-\to \Xi_c^{\prime 0} \bar \Lambda_c^{-}$ rate. 
Other modes have large uncertainties in their rates.
This simply reflect our poor understanding of the SU(3) breaking parameters $\delta^{c_2}_v$, $\delta^{c_2}_c$ and $\delta^{c_2}_s$.
Hence, measurements of these rates can provide valuable information about these SU(3) breaking parameters.

\subsubsection{Excited $\B_c({\bf 6_f})$ case}

We now turn to the numerical results of $\overline B_q\to \B_c({\bf 6_f})\overline \B_c ({\bf 3_f})$ decays for low-lying $\overline \B_c ({\bf 3_f})$ charmed baryons, but with excited $\B_c({\bf 6_f})$ anti-charmed baryons.
Namely, we will discuss
$\overline B_q\to \Sigma_c^{0,+} (2520)\overline \B_c ({\bf 3_f})$, $\Xi_c^{0,+}(2645)\overline \B_c ({\bf 3_f})$ and $\Omega_c^{0}(2770)\overline \B_c ({\bf 3_f})$ decays.
As shown in Table~\ref{tab: quantum number}, these charmed baryons are $J^P=\frac{3}{2}^+$ members of a ${\bf 6_f}$ multiplet.
Note that, as forbidden by charge conservation, we do not have the $\overline B_q\to \Sigma_c^{++} (2520)\overline \B_c ({\bf 3_f})$ decay.
The amplitudes of these modes are governed by internal $W$-tree amplitudes, and can be expressed similarly as those in Table~\ref{tab: BtoBcBcbar II}.
For example, we have 
\be
A(B^-\to \Xi_c^{0}(2645) \bar \Lambda_c^{-}=-\tilde C_{2,v},
\en
where the tilde on top of the topological amplitude is to differentiate from the amplitudes of the low-lying case.

\begin{table}[t!]
\caption{\label{tab: cp2}
Parameters in $\overline B_q\to \B_c({\bf 6_f})\overline \B_c ({\bf 3_f})$ for low lying $\overline \B_c ({\bf 3_f})$ and excited $(3/2)^+$ $\B_c({\bf 6_f})$ states, 
where $\tilde c_{2,v}\equiv (1+\delta^{\tilde c_2}_v) \tilde c_2$ is a fitted parameter. 
The uncertainty is obtained by scanning $\chi^2\leq \chi^2_{\rm min.}+1$.
We assume SU(3) breaking parameters ($\delta^{c_2}_v$, $\delta^{c_2}_c$, $\delta^{c_2}_s$) have sizes at most as $|\delta^{c_1}_v|$.}
\begin{ruledtabular}
\begin{tabular}{ccccc} 
 $\tilde c_{2,v} \equiv (1+\delta^{\tilde c_2}_v) \tilde c_2$
          & $\delta^{\tilde c_2}_v$ 
          & $\delta^{\tilde c_2}_c$
          & $\delta^{\tilde c_2}_s$
          & $\chi^2_{\rm min.}$
         \\
          \hline
$0.579_{-0.189}^{+0.141}$
          & $0\pm 0.35$
          & $0\pm 0.35$
          & $0\pm 0.35$
          & 0
          \\
\end{tabular}
\end{ruledtabular}
\end{table}

\begin{table}[t!]
\caption{\label{tab: br BtoBcBcbar 6p3}
$\overline B_q\to \B_c({\bf 6_f})\overline \B_c ({\bf 3_f})$ decay amplitudes and branching ratios (in unit of $10^{-4}$) in $\Delta S=-1$ and  $\Delta S=0$ transitions,
where $\B_c ({\bf \bar 3_f})$ are low lying $(1/2)^+$ states, while $\B_c({\bf 6_f})$ are excited $(3/2)^+$ states .
The experimental result is the input of the $\chi^2$-fit.}
\footnotesize{
\begin{ruledtabular}
\begin{tabular}{lccc}
Mode
          & $A \left(\overline B_q\to \B_c({\bf 6_f})\overline \B_c ({\bf 3_f})\right)$
          & $Br \left(\overline B_q\to \B_c({\bf 6_f})\overline \B_c ({\bf 3_f})\right)$
          & Expt
          \\
\hline
           $B^-\to \Xi_c^{0}(2645) \bar \Lambda_c^{-}$
          & $-\tilde C_{2,v}$
          & $4.40\pm 2.40$
          & $4.4\pm 2.4$ \cite{Belle:2019pze} 
          \\
$\overline B{}^0\to \Xi_c^{+}(2645) \bar \Lambda_c^{-}$
          & $\tilde C_{2,v}$
          & $4.09\pm 2.23$
          & $-$ 
          \\
$\overline B{}_s^0\to \Xi_c^{+}(2645) \bar \Xi_c^{-}$
          & $-\tilde C_{2,vs}$
          & $2.67_{-2.41}^{+7.08}$
          & $-$ 
          \\
$\overline B{}_s^0\to \Xi_c^{0}(2645) \bar \Xi_c^{0}$
          & $\tilde C_{2,vs}$
          & $2.61_{-2.36}^{+6.89}$
          & $-$ 
           \\
$B^-\to \Omega_c^{0}(2770) \bar \Xi_c^{-}$
          & $\sqrt2 \tilde C_{2,vc}$
          & $0.40_{-0.36}^{+1.07}$
          & $-$ 
          \\    
$\overline B{}^0\to \Omega_c^{0}(2770) \bar \Xi_c^{0}$
          & $-\sqrt2 \tilde C_{2,vc}$
          & $0.34_{-0.31}^{+0.91}$
          & $-$ 
          \\
\hline
         $B^-\to \Sigma_c^0 (2520) \bar \Lambda_c^{-}$
          & $-\sqrt2 \tilde C'_2$
          & $0.81_{-0.61}^{+2.17}$
          & $-$ 
          \\
$\overline B{}^0\to \Sigma_c^+ (2520)\bar \Lambda_c^{-}$
          & $\tilde C'_2$
          & $0.38_{-0.28}^{+1.01}$
          & $-$ 
          \\                  
$\overline B{}_s^0\to \Sigma_c^+ (2520)\bar \Xi_c^{-}$
          & $-\tilde C'_{2,s}$
          & $0.28_{-0.25}^{+1.61}$
          & $-$ 
          \\
$\overline B{}_s^0\to \Sigma_c^0 (2520)\bar \Xi_c^{0}$
          & $\sqrt2 \tilde C'_{2,s}$
          & $0.56_{-0.50}^{+3.17}$
          & $-$ 
          \\                                            
$B^-\to \Xi_c^{0} (2645) \bar \Xi_c^{-}$
          & $\tilde C'_{2,c}$
          & $0.08_{-0.07}^{+0.45}$
          & $-$ 
         \\
$\overline B{}^0\to \Xi_c^{0} (2645) \bar \Xi_c^{0}$
          & $-\tilde C'_{2,c}$ 
          & $0.07_{-0.06}^{+0.41}$
          & $-$ 
          \\
\end{tabular}
\end{ruledtabular}
}
\end{table}

On the experimental side, so far, the measurement of the $B^-\to \Xi_c^{0}(2645) \bar \Lambda_c^{-}$ rate among these decays is reported, see Table~\ref{tab: expt}.
Hence, only information about $|\tilde C_{2,v}|$, or equivalently $|\tilde c_{2,v}|= |(1+\delta^{\tilde c_2}_v) \tilde c_2|$ can be obtained.
The fitted result of this parameter is shown in Table~\ref{tab: cp2}.
By assuming SU(3) breaking parameters ($\delta^{\tilde c_2}_v$, $\delta^{\tilde c_2}_c$, $\delta^{\tilde c_2}_s$) have sizes at most as $|\delta^{c_1}_v|$,
results of other $\overline B_q\to \B_c({\bf 6_f})\overline \B_c ({\bf 3_f})$ decay rates can be obtained and are shown in Table~\ref{tab: br BtoBcBcbar 6p3}.
We also show their amplitudes for our convenience.

Note that as these are $\overline B_q\to \B_c(J=\frac{3}{2})\overline \B_c (J=\frac{1}{2})$ decays, the decay rates are proportional to their center of mass momenta cubed, $p_{cm}^3$, see Eq. (\ref{eq: rate 3/2 1/2}).
These factors provide large suppressions on rates for heavy final states.
For example, the $B^-\to \Omega_c^{0}(2770) \bar \Xi_c^{-}$ and $\overline B{}^0\to \Omega_c^{0}(2770) \bar \Xi_c^{0}$ rates are much smaller than other $\Delta S=-1$ modes.
This is even more prominent by comparing this situation to those of $B^-\to \Omega_c^{0} \bar \Xi_c^{-}$ and $\overline B{}^0\to \Omega_c^{0} \bar \Xi_c^{0}$ rates given in Table~\ref{tab: br BtoBcBcbar 63}, where the suppression is much milder as they are $\overline B_q\to \B_c(J=\frac{1}{2})\overline \B_c (J=\frac{1}{2})$ decays instead.

Except for $\overline B{}^0\to \Xi_c^{+}(2645) \bar \Lambda_c^{-}$, 
which is the isospin related mode of $B^-\to \Xi_c^{0}(2645) \bar \Lambda_c^{-}$, all other modes have large uncertainties in their rates.
This simply reflect our poor understanding of the SU(3) breaking parameters $\delta^{\tilde c_2}_v$, $\delta^{\tilde c_2}_c$ and $\delta^{\tilde c_2}_s$.
Hence, measurements of these rates can provide valuable information about these SU(3) breaking parameters.

\subsection{$\overline B_q\to \B_c({\bf \bar 3_f})\overline \B_c ({\bf \bar 6_f})$ decay rates}

Finally we discuss the numerical results of $\overline B_q\to \B_c({\bf \bar 3_f})\overline \B_c ({\bf \bar 6_f})$ decay rates with low-lying charmed and anti-charmed baryons.
These modes are governed by internal $W$-tree amplitudes as shown in Table~\ref{tab: BtoBcBcbar III}.
Presently, we have measurements on the rates of two of these modes, 
namely $B^-\to \Xi_c^+ \bar \Sigma_c^{--}$ and $\overline B{}^0\to \Xi_c^0 \bar \Sigma_c^{0}$ decays, 
see Table~\ref{tab: expt}. These two modes are isospin-related and are governed by $C_{3,v}$.
Hence, we have information on $|C_{3,v}|$, or equivalently $|c_{3,v}|= |(1+\delta^{c_3}_v) c_3|$.
The fitted result of this parameter is shown in Table~\ref{tab: c3}.

\begin{table}[t!]
\caption{\label{tab: c3}
Parameters in $\overline B_q\to \B_c({\bf \bar 3_f})\overline \B_c ({\bf \bar 6_f})$ decays for low lying $\B_c({\bf \bar 3_f})$ and $\overline \B_c ({\bf \bar 6_f})$, 
where $c_{3,v}\equiv (1+\delta^{c_3}_v) c_3$ is a fitted parameter. 
The uncertainty is obtained by scanning $\chi^2\leq \chi^2_{\rm min.}+1$.
We assume SU(3) breaking parameters ($\delta^{c_3}_v$, $\delta^{c_3}_c$, $\delta^{c_3}_s$) have sizes at most as $|\delta^{c_1}_v|$.}
\begin{ruledtabular}
\begin{tabular}{ccccc} 
 $c_{3,v} \equiv (1+\delta^{c_2}_v) c_3$
          & $\delta^{c_3}_v$ 
          & $\delta^{c_3}_c$
          & $\delta^{c_3}_s$
          & $\chi^2_{\rm min}$
         \\
          \hline
$0.159_{-0.020}^{+0.017}$
          & $0\pm 0.35$
          & $0\pm 0.35$
          & $0\pm 0.35$
          & $0.035$
          \\
\end{tabular}
\end{ruledtabular}
\end{table}

\begin{table}[t!]
\caption{\label{tab: Br BtoBcBcbar 3bar6bar}
Branching ratios (in unit of $10^{-4}$) of $\overline B_q\to \B_c({\bf \bar 3_f})\overline \B_c ({\bf \bar 6_f})$ decays to low-lying charmed baryon final states in $\Delta S=-1$ and  $\Delta S=0$ transitions.
Experimental results are inputs of the $\chi^2$-fit giving $\chi^2_{\rm min.}=0.035$.}

\footnotesize{
\begin{ruledtabular}
\begin{tabular}{lccc}
Mode
          & $A \left(\overline B_q\to \B_c({\bf \bar 3_f})\overline \B_c ({\bf \bar 6_f})\right)$
          & $Br\left(\overline B_q\to \B_c({\bf \bar 3_f})\overline \B_c ({\bf \bar 6_f})\right)$
          & Expt
          \\
\hline
          $B^-\to \Xi_c^+ \bar \Sigma_c^{--}$
          & $\sqrt2 C_{3,v}$
          & $5.38\pm1.24$
          & $5.74\pm2.33$ \cite{Belle:2025nup}
          \\ 
$\overline B{}^0\to \Xi_c^0 \bar \Sigma_c^{0}$
          & $-\sqrt2 C_{3,v}$
          & $4.96\pm 1.14$
          & $4.83\pm1.35$ \cite{Belle:2025nup}
          \\
$B^-\to \Xi_c^0 \bar \Sigma_c^{-}$
          & $-C_{3,v}$ 
          & $2.68_{-0.62}^{+0.62}$
          & $-$
          \\     
$\overline B{}^0\to \Xi_c^+ \bar \Sigma_c^{-}$
          & $C_{3,v}$
          & $2.49\pm 0.57$
          & $-$
          \\
$\overline B{}_s^0\to \Xi_c^0 \bar \Xi_c^{\prime 0}$
          & $-C_{3,vs}$
          & $2.37_{-1.98}^{+4.53}$
          & $-$
          \\
$\overline B{}_s^0\to \Xi_c^+ \bar \Xi_c^{\prime -}$
          & $C_{3,vs}$
          & $2.38_{-1.99}^{+4.56}$
          & $-$
          \\
\hline
$B^-\to \Lambda_c^+ \bar \Sigma_c^{- -}$
          & $-\sqrt2 C'_3$
          & $0.35_{-0.20}^{+0.67}$
          & $-$
          \\
$\overline B{}^0\to \Lambda_c^+ \bar \Sigma_c^{-}$
          & $-C'_3$
          & $0.16_{-0.09}^{+0.31}$
          & $-$
          \\
$B^-\to \Xi_c^0 \bar \Xi_c^{\prime -}$
          & $C'_{3,c}$
          & $0.12_{-0.10}^{+0.50}$
          & $-$
          \\
$\overline B{}^0\to \Xi_c^0 \bar \Xi_c^{\prime 0}$
          & $C'_{3,c}$
          & $0.11_{-0.09}^{+0.46}$
          & $-$ 
          \\          
$\overline B{}^0_s\to \Lambda_c^+ \bar \Xi_c^{\prime -}$
          & $-C'_{3,s}$
          & $0.16_{-0.13}^{+0.68}$
          & $-$
          \\         
$\overline B{}_s^0\to \Xi_c^0 \bar \Omega_c^{0}$
          & $\sqrt2 C'_{3,cs}$
          & $0.20_{-0.19}^{+1.50}$
          & $-$
          \\                                            
\end{tabular}
\end{ruledtabular}
}
\end{table}

By assuming SU(3) breaking parameters $\delta^{c_3}_v$, $\delta^{c_3}_c$ and $\delta^{c_3}_s$ have sizes at most as $|\delta^{c_1}_v|$, rates of these modes can be obtained and are shown in Table~\ref{tab: Br BtoBcBcbar 3bar6bar}.
We see that the experimental data on $B^-\to \Xi_c^+ \bar \Sigma_c^{--}$ and $\overline B{}^0\to \Xi_c^0 \bar \Sigma_c^{0}$ decay rates can be reasonably reproduced.
Note that in the fit, the central value of the $B^-\to \Xi_c^+ \bar \Sigma_c^{--}$ rate is slightly lower than that in data, while the situation in $\overline B{}^0\to \Xi_c^0 \bar \Sigma_c^{0}$ decay is the other way around. Hence, the minimum of $|C_{3,v}|$ is constrained by the former mode, while the maximum is by the latter mode.

The rates of two other isospin-related modes, namely $B^-\to \Xi_c^0 \bar \Sigma_c^{-}$ and $\overline B{}^0\to \Xi_c^+ \bar \Sigma_c^{-}$ can be predicted with relatively small uncertainties.
The uncertainties in rates of other modes are much larger.
This simply reflect our poor understanding of the SU(3) breaking parameters $\delta^{c_3}_v$, $\delta^{c_3}_c$ and $\delta^{c_3}_s$.
Hence, measurements of these rates can provide valuable information about these SU(3) breaking parameters.

\section{Conclusion}

We study the rates of two-body charmed anti-charmed baryonic $\overline B\to {\cal B}_c \overline {\cal B}_c$ decays using the topological amplitude approach.
All amplitudes of $\overline B\to {\cal B}_c(\bf {\bar 3_f}) \overline {\cal B}_c(\bf { 3_f})$,
${\cal B}_c(\bf 6_f) \overline {\cal B}_c(\bf { 3_f})$,
${\cal B}_c(\bf {\bar 3_f}) \overline {\cal B}_c(\bf {\bar 6_f})$
and
${\cal B}_c(\bf 6_f) \overline {\cal B}_c(\bf {\bar 6_f})$ decays
are decomposed topologically.
SU(3) breaking effects on these amplitudes, depending on the position of the $s$-quark line, are modeled.
Note that the numerical results of this work rely on several working assumptions. They can be relaxed when more data becomes available. For example, our estimate of some SU(3) breaking parameters can be relaxed, and the relative contributions from $s$-wave and $p$-wave parts can be determined with the information of the asymmetry $\alpha$.

Using existing data as inputs, we obtained the following results. 
\begin{itemize}
\item[(i)] In the low-lying $\overline B\to {\cal B}_c(\bf {\bar 3_f}) \overline {\cal B}_c(\bf { 3_f})$ decays, 
experimental results on $B^-\to \Xi_c^0 \bar \Lambda_c^{-}$,
$\overline B{}^0\to \Xi_c^+ \bar \Lambda_c^{-}$,
$\overline B{}_s^0\to \Lambda_c^+ \bar \Lambda_c^{-}$
and $\overline B{}^0\to \Lambda_c^+ \bar \Lambda_c^{-}$ decay rates
can be reasonably reproduced, and predictions on other rates are given.
We find that
the exchange diagram is sizable, giving  $|E^{(\prime)}_1|/|C^{(\prime)}_1|$ ratio about 44\%.
Furthermore, there is a large cancellation in internal $W$-tree and exchange $W$-tree amplitudes.   
The SU(3) breaking is sizable, 35\% SU(3) breaking effects are needed, and they work differently in different amplitudes.
Namely, the internal $W$-tree amplitude $C_{1,v}$ is enlarged, 
while the exchange-$W$ tree amplitude $E_{1,v}$ is reduced.
It remains to be seen if this pattern is followed by other internal $W$-tree and exchange-$W$ tree amplitudes,
such as $C_{1,s}$, $C_{1,c}$, $C_{1,vs}$, $E_{1,s}$ and $E_{1,vc}$.
The rates of $\overline B\to {\cal B}_c(\bf {\bar 3_f}) \overline {\cal B}_c(\bf { 3_f})$ decays with excited ${\cal B}_c(\bf {\bar 3_f})$, such as 
$\Lambda_c(2595)^+$, $\Xi_c(2790)^{+,0}$, are also studied.
The experimental result on the $B^-\to\Xi_c^{0}(2790)\bar \Lambda_c^-$ rate can be reproduced with predictions on rates of some similar modes.
Note that some of the above observations on the low-lying case have been pointed out in \cite{LHCb:2025ueu}, 
but, in this work, we provide more detailed information, 
such as the size of the ratio of the topological amplitudes, 
the sizes of the SU(3) breaking effects, and different behaviors of the effects on different topological amplitudes.

\item[(ii)] The $\overline B\to {\cal B}_c(\bf 6_f) \overline {\cal B}_c(\bf { 3_f})$ decays, 
with low-lying $ \overline {\cal B}_c(\bf { 3_f})$ and low-lying and excited ${\cal B}_c(\bf 6_f)$ baryons, 
such as $\Sigma_c(2520)^{+,0}$, $\Xi_c(2645)^{+,0}$, $\Omega_c(2770)^0$, are studied.
Since these decays are governed by a single internal $W$-tree amplitude in the SU(3) limit, they are highly related.
Adding SU(3) breaking can easily modify their relations.
Experimental results on $B^-\to \Xi_c^{\prime 0} \bar \Lambda_c^{-}$ and $B^-\to \Xi_c^{0}(2645) \bar \Lambda_c^{-}$ decays can be reproduced with predictions on other rates obtained. 
Note that the excited states are spin-3/2 states. Therefore, some of their rates are highly suppressed by the kinematic factor ($p_{cm}^3$).

\item[(iii)] The $\overline B\to {\cal B}_c(\bf {\bar 3_f}) \overline {\cal B}_c(\bf {\bar 6_f})$ decays with low-lying charmed anti-charmed baryons are studied.
Since these decays are governed by a single internal $W$-tree amplitude in the SU(3) limit, they are highly related.
Adding SU(3) breaking can easily modify their relations.
The experimental data on $B^-\to \Xi_c^+ \bar \Sigma_c^{--}$ and $\overline B{}^0\to \Xi_c^0 \bar \Sigma_c^{0}$ decay rates can be reasonably reproduced
with predictions on other rates given.

\item[(iv)] Uncertainties in most predicted rates are large, reflecting our current poor understanding of the related SU(3) breaking effects.
Measuring these rates can provide very useful information about these effects.

\end{itemize}

\begin{acknowledgments}
This work is supported in part by the National Science and Technology Council of R.O.C.
under Grant No NSTC-114-2112-M-033-002.
\end{acknowledgments}

\appendix

\section{Formula for decay rates}\label{App: Br}

The decay $\overline  B_q\to \B_c(J=1/2) \bar\B_c(J=1/2)$ and $ \B_c(J=3/2) \bar\B_c(J=1/2)$ decay have the following forms~\cite{Jarfi:1990ej}
\begin{eqnarray}
A\left(\overline B_q\to \B_c\left(\frac{1}{2}\right) \bar\B_c\left(\frac{1}{2}\right)\right)
&=&\bar u_1(a+\gamma_5 b) v_2,
\nonumber\\
A\left(\overline B_q\to \B_c\left(\frac{3}{2}\right) \bar\B_c\left(\frac{1}{2}\right)\right)
&=&
i \frac{q^\mu}{m_{B_q}} \bar u^\mu_1(a+\gamma_5 b) v_2,
\label{eq: A1}
\end{eqnarray}
where $q=p_1-p_2$ is the difference of the momenta of the charmed and anti-charmed baryons and $u^\mu,\,v^\mu$ are the Rarita-Schwinger vector spinors,~\cite{Moroi:1995fs}
\be
u_\mu\left(\pm\frac{3}{2}\right)&=&\epsilon_\mu(\pm1) u\left(\pm\frac{1}{2}\right)
\non\\
u_\mu\left(\pm\frac{1}{2}\right)&=&
\frac{1}{\sqrt3}\left(\epsilon_\mu(\pm1)
u\left(\mp\frac{1}{2}\right)+\sqrt{2}\,\epsilon_\mu(0)
u\left(\pm\frac{1}{2}\right)\right),
\en 
with $\epsilon_\mu(\lambda)$ the polarization vector.
Using the following identity
\be
q\cdot\epsilon(\lambda)&=&
-\,\delta_{\lambda,0}\, \frac{ m_{B_q} p_{cm}}{m_{\B_c(3/2)}},
\en
with $p_{cm}$ the baryon momentum in the center of mass frame, we have
\begin{eqnarray}
A\left(\overline B_q\to \B_c\left(\frac{3}{2}\right) \bar\B_c\left(\frac{1}{2}\right)\right)
&=&-i
\sqrt{\frac{2}{3}}\frac{p_{cm}}{m_{\B_c(3/2)}} \bar u_1(a+\gamma_5 b) v_2.
\label{eq: 3/2 1/2}
\end{eqnarray}
 
 It is straightforward to obtain the decay rates giving
\be
\Gamma\left(\overline B_q\to \B_c\left(\frac{1}{2}\right) \bar\B_c\left(\frac{1}{2}\right)\right)
&=&
\frac{p_{cm}}{8 \pi m_{B_q}^2} [(2 m_{B_q}^2 - 2 (m_{\B_c}+ m_{\bar\B_c})^2) |a|^2
\non\\
&&\qquad
+(2 m_{B_q}^2 - 2 (m_{\B_c}- m_{\bar\B_c})^2) |b|^2],
\en
and
\be
\Gamma\left(\overline B_q\to \B_c\left(\frac{3}{2}\right) \bar\B_c\left(\frac{1}{2}\right)\right)
&=&
\frac{p_{cm}}{8 \pi m_{B_q}^2}
\frac{2}{3}
\left(\frac{p_{cm}}{m_{\B_c(3/2)}} \right)^2
[ (2 m_{B_q}^2 - 2 (m_{\B_c}+ m_{\bar\B_c})^2) |a|^2
\non\\
&&\qquad
+(2 m_{B_q}^2 - 2 (m_{\B_c}- m_{\bar\B_c})^2) |b|^2].
\en


\end{document}